\documentclass[sigconf]{acmart} 

\usepackage{booktabs}

\usepackage[ruled]{algorithm2e} 
\SetAlFnt{\small}
\SetAlCapFnt{\small}
\SetAlCapNameFnt{\small}
\SetAlCapHSkip{0pt}

\usepackage{fancyvrb}

\usepackage{mathtools}

\usepackage{multirow}

\usepackage{wrapfig}

\usepackage{subcaption}

\copyrightyear{2025}
\acmYear{2025}
\setcopyright{cc}
\setcctype{by}
\acmConference[UIST '25]{The 38th Annual ACM Symposium on User Interface Software and Technology}{September 28-October 1, 2025}{Busan, Republic of Korea}
\acmBooktitle{The 38th Annual ACM Symposium on User Interface Software and Technology (UIST '25), September 28-October 1, 2025, Busan, Republic of Korea}
\acmDOI{10.1145/3746059.3747751}
\acmISBN{979-8-4007-2037-6/2025/09}

\newcommand{\sba}{\raisebox{0.005in}[0pt]{\includegraphics[width=0.15in]{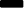}}}

\newcommand{\sbc}{\raisebox{-0.030in}[0pt]{\includegraphics[width=0.15in]{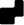}}}

\newcommand{\refAppendixVersionTwoA}
  {\autoref{sec:version-2a}~\cite{Rocks}}
\newcommand{\refAppendixVersionTwoB}
  {\autoref{sec:version-2b}~\cite{Rocks}}
\newcommand{\refAppendixApplications}
  {\autoref{sec:applications}~\cite{Rocks}}
\newcommand{\refAppendixTimetableAbs}
  {\autoref{sec:timetable-abs}~\cite{Rocks}}
\newcommand{\refAppendixMeshDistance}
  {\autoref{sec:mesh-distance}~\cite{Rocks}}

\newcommand{\ie}{i.e.,}
\newcommand{\eg}{e.g.,}

\newcommand{\List}[1]{[#1]}

\newcommand{\Constr}[2]{\textsf{#1}(#2)}
\newcommand{\Unit}[1]{\textsf{#1}}

\newcommand{\Wrap}[2]{\textsf{Wrap}_{#1}(#2)}
\newcommand{\Node}[2]{\textsf{Node}_{#1}(#2)}
\newcommand{\Cell}[2]{\textsf{Cell}_{#1}(#2)}
\newcommand{\Stack}[1]{\Constr{Stack}{#1}}
\newcommand{\Pin}[2]{\textsf{Pin}_{#1}(#2)}

\newcommand{\term}[1]{\textit{#1}}
\newcommand{\pdn}{$\raisebox{0.007in}{::}{=}$}

\newcommand{\fn}[1]{\textrm{#1}}

\newcommand{\Algop}[1]{${\textsc{#1}}_\text{P}$}
\newcommand{\Algos}[1]{${\textsc{#1}}_\text{S}$}
\newcommand{\Algoss}[1]{${\textsc{#1}}_\text{S}+\text{S}$}
\newcommand{\Algoi}{${\textsc{L1}}^\textit{i}_\text{P}$}
\newcommand{\Boxes}{\textsc{Boxes}}
\newcommand{\Boxesns}{$\textsc{Boxes}_{\text{NS}}$}
\newcommand{\SBlock}{\textsc{S-Blocks}}

\newcommand{\Abs}{\textsc{Abs}}

\newcommand{\joinv}{\ensuremath{\fn{join}_\text{V}}}
\newcommand{\joinh}{\ensuremath{\fn{join}_\text{H}}}
\newcommand{\unionr}{\ensuremath{\fn{union}_\text{R}}}
\newcommand{\unionl}{\ensuremath{\fn{union}_\text{L}}}
\newcommand{\wraps}{\ensuremath{\fn{wrap}_\text{S}}}
\newcommand{\wrapr}{\ensuremath{\fn{wrap}_\text{R}}}
\newcommand{\wrapl}{\ensuremath{\fn{wrap}_\text{L}}}

\definecolor{newcolor}{RGB}{213 245 175}
\newcommand{\new}[1]{\colorbox{newcolor}{#1}}
\definecolor{changedcolor}{RGB}{255 251 210}
\newcommand{\changed}[1]{\colorbox{changedcolor}{#1}}
\definecolor{removedcolor}{RGB}{255 204 203}
\newcommand{\removed}[1]{\colorbox{removedcolor}{#1}}

\newcommand{\letter}[1]{\raisebox{-0.023in}{\includegraphics{images/letters/#1.pdf}}}

\newcommand{\rocks}{\textsc{Rocks}}
 
\begin{document}
\title[Ragged Blocks]{Ragged Blocks: Rendering Structured Text with Style}

\settopmatter{authorsperrow=2}

\author{Sam Cohen}
\orcid{0009-0008-9127-6518}
\email{samcohen@uchicago.edu}
\affiliation{\institution{University of Chicago}
\city{Chicago}
  \state{IL}
  \country{USA}
  \postcode{60637}
}
\author{Ravi Chugh}
\orcid{0000-0002-1339-2889}
\email{rchugh@cs.uchicago.edu}
\affiliation{\institution{University of Chicago}
\city{Chicago}
  \state{IL}
  \country{USA}
  \postcode{60637}
}

\begin{abstract}
  Whether it be source code in a programming language,
  prose in natural language, or otherwise,
  text is highly structured.
Currently, text visualizations are confined either to \emph{flat, line-based} decorations, which can convey only limited information about textual structure, or \emph{nested boxes}, which convey structure but often destroy the typographic layout of the underlying text.
We hypothesize that the lack of rich styling options limits the kinds of information that are displayed alongside text, wherever it may be displayed.

  In this paper, we show that it is possible to achieve arbitrarily nested decorations while minimally disturbing the underlying typographic layout.
Specifically, we present a layout algorithm that generates \emph{ragged blocks}, or \emph{rocks}, which are rectilinear polygons that allow nested text to be compactly rendered even when styled with borders and padding.
Our layout algorithm is evaluated
on a benchmark suite comprising representative source code files in multiple programming languages. The (ragged block) layouts produced by our algorithm are substantially more compact than the (rectangular block) layouts produced by conventional techniques, when uniformly styling every element in the syntax tree with borders and padding.
\end{abstract}

\ccsdesc[300]{Human-centered computing~Graphical user interfaces}
\ccsdesc[300]{Human-centered computing~Visualization}
\ccsdesc[300]{Software and its engineering~Integrated and visual development environments}
\ccsdesc[300]{Computing methodologies~Computer graphics}

\keywords{Structured Text,
Text Layout,
Ragged Blocks,
Program Visualization
}

\begin{teaserfigure}
  \includegraphics[width=7in]{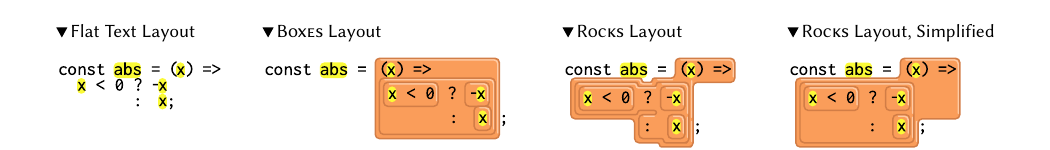}
  \caption{A code fragment rendered without nested styling, with nested boxes,
with rocks, and with simplified rocks.}
  \vspace{0.25in}
\label{fig:blocks-vs-rocks}
\end{teaserfigure}
 
\maketitle

\begin{figure*}
  \includegraphics[width=7in]{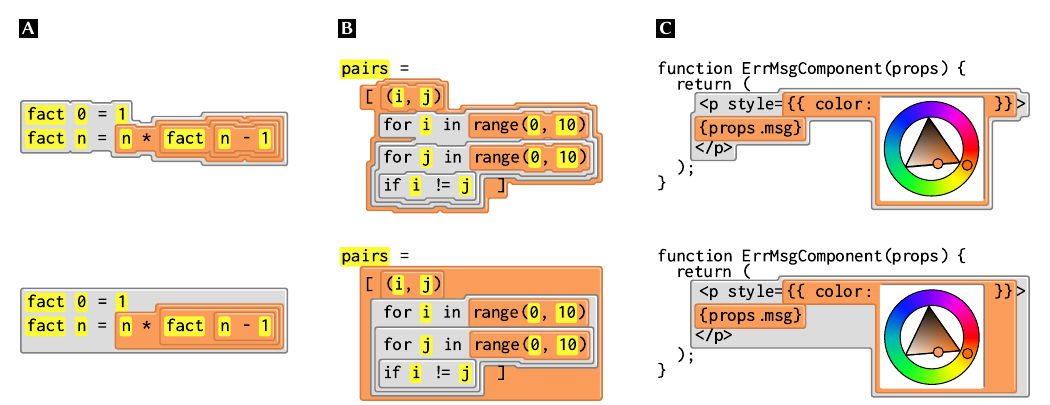}
  \caption{Factorial, List Comprehension, and Color Picker. Each is shown rendered as rocks, with and without simplification.}
  \label{fig:intro}
\end{figure*}
 
\section{Introduction}
\label{sec:introduction}

Document-rendering GUIs currently do not provide expressive mechanisms for styling and rendering structured text.

In rich text editors and word processors, such as Google Docs and Microsoft Word, document authors call attention to parts of their documents using \emph{text decorations}.
These decorations include underlines, italics, foreground and background colors, and inline glyphs, among others.
When communicating information back to users, these tools also use such decorations; for example, wavy underlines are commonly used to identify \raisebox{-0.04in}[0pt]{\includegraphics{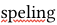}} errors and opportunities to make grammar \raisebox{-0.04in}[0pt]{\includegraphics{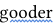}}.
These decorations are \emph{flat}: they can appear only within or between lines of text, and are thus inherently limited in their capacity to identify structures of text besides tokens and lines.

In traditional code editors and IDEs, such as CodeMirror and VS Code, program authors do not directly apply styles to the source text.
But these interfaces typically render programs with syntax highlighting---where tokens are colored based on lexical and ``parts-of-speech''--type information---and employ flat decorations, such as wavy underlines and simple glyphs, to identify errors.

In contrast to traditional code editors, \emph{block-based editors}, such as Scratch~\cite{Resnick2009} and Snap!~\cite{Snap}, employ shapes that allow certain code fragments to be ``snapped'' together via mouse-based direct manipulation.
Other \emph{structure editors}, such as Sandblocks~\cite{Sandblocks2020, Sandblocks2023}, render rectangular blocks (\ie{}~boxes) for all syntactic substructures in an otherwise text-based, keyboard-driven editor.

Sadly, current text layout algorithms for all of the above systems compromise too much in order to surface structure.
They are typically line-based, where visual indications of structure cannot cross the line boundary, or block-based, which often force the underlying text into a layout which does not resemble its source.
The ``\Boxes{} Layout'' in \autoref{fig:blocks-vs-rocks} shows an example deficiency of the latter approach:
notice how the nested rectangles, like many GUI elements found in visual programming interfaces, ``occupy a great deal of space, making it harder to maintain context''~\cite{BurnettVP}.
Furthermore, we hypothesize that layouts which minimize the \emph{error} with respect to plain-text displays will help facilitate a workflow---described in our prior work on \emph{code style sheets}~\cite{Hass}---where programs and other structured text can be styled and restyled in flexible ways.

\paragraph{Ragged Blocks}

The idea driving this work is to use \emph{ragged blocks}---abbreviated to \emph{rocks}---rather than rectangles to render nested substructures of text for rich styling.
Analogous to how ``ragged right'' text is aligned on the left margin but unaligned on the right, a rock is an arbitrary rectilinear polygon, which is ``ragged on all sides.''

\autoref{fig:intro} demonstrates three code snippets rendered as ragged blocks:
\letter{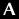}~a factorial function written in Haskell,
\letter{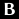}~a list comprehension with multiple iterables in Python, and
\letter{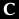}~an error message component in TypeScript.
The top row of this figure shows \emph{compact} rock layouts, where rocks for different substructures have sometimes highly irregular boundaries in order to tightly fit the text contained within.
Once tight-fitting rock layouts have been computed, it is possible to derive \emph{simplified} rock layouts (the second row of \autoref{fig:intro}), where some corners have been removed (\ie{}~some edges have been straightened) without further affecting the position of laid-out elements.

Although our primary motivation is text, the notion of ragged-block layout is agnostic to the contents of individual elements.
For example, \autoref{fig:intro}~\letter{c} shows a color-picking GUI compactly embedded within the surrounding program text.
\emph{Projectional editors}, such as Hazel~\cite{OmarLivelits} and MPS~\cite{MPS}, enrich code editors with GUIs in this way, but---like block-based and structure editors---these systems do not render nested blocks as compactly as shown in \autoref{fig:intro}.

\paragraph{Contributions}

The primary contribution of this paper is
a family of layout algorithms that compactly render highly structured text documents as nested ragged blocks.
The key challenge is the presence of newlines in the source text.
Whereas conventional nested-box layout algorithms limit the visual effect of a newline to the box containing it, a ragged-block layout algorithm must visually reconcile arbitrarily large subtrees separated by a newline.
We formulate a data structure, called a \term{region}, that reifies the layered steps in our layout algorithm in a way that allows cross-line visual constraints to be permeated and reconciled efficiently.

We evaluate our principal layout algorithm on a benchmark suite comprising 7.6k lines of code written in Haskell, Python, and TypeScript, by comparing the deviation of rock- and conventional nested-box layouts to unstyled text.
For this evaluation, we uniformly render all substructures in the syntax tree with border and padding; these ``blocks-everywhere'' visualizations serve as a stress test for nested text layout.

\paragraph{Outline}

The rest of this paper is organized is follows.
In \autoref{sec:related-work}, we discuss related work with a focus on program visualization since it is a particularly rich source of examples for structured text layout.

In \autoref{sec:version-1}, we motivate and present the {region} data structure, which underlies our family of ragged block layout algorithms.
We describe the simplest version of our layout algorithm, called \Algop{L1}, in detail. The label ``P'' indicates the use of \emph{pure} regions in this algorithm.

The next section presents several extensions to the base algorithm, \Algop{L1}.
In \autoref{sec:extensions-2a} and \autoref{sec:extensions-2b}, we describe two modest variations, called \Algop{L2a} and \Algop{L2b}, which provide more nuanced treatment of whitespace characters for code and prose, respectively.
In \autoref{sec:stateful-regions}, we present an alternative representation of regions, called \emph{stateful} regions, which enables sharing among regions and enables performance optimizations; the correspondingly revised algorithms, called \Algos{L1}, \Algos{L2a}, and \Algos{L2b}, respectively, are labeled ``S.''

Each of these layout algorithms computes tightly-packed ragged blocks.
In \autoref{sec:simplification}, we define an optional post-processing algorithm, called \emph{simplification}, that makes borders less ragged where possible without affecting the positions of text fragments in the layout.

To evaluate our principal layout algorithm, \Algos{L1},
in \autoref{sec:benchmarks} we devise a metric for measuring the \emph{distance} between two text layouts.
Our experiments show that \Algos{L1} produces layouts that, when considering plain-text layouts as the point of reference, have less error than those produced by a conventional nested-box algorithm as well as a recently proposed algorithm in the literature.
We conclude in \autoref{sec:disco} with a discussion of potential user interface applications of rocks layout and other avenues for future work.\footnote{In the rest of the paper, we write \rocks{} (typeset in small caps) to refer collectively to our implementations of the layout algorithms, together with a simple parsing pipeline that allows us to generate the examples depicted throughout. Our implementation is available at \url{https://github.com/sbcohen2000/ragged-blocks/}, and additional technical details can be found in an extended report~\cite{Rocks}.}

\section{Related Work}
\label{sec:related-work}

Structured text abounds.
How is it currently rendered?

\paragraph{Code Editors}
There is a long history of work on \emph{block-based}, \emph{structure}, and \emph{projectional} code editors that blend source code text and visual elements in myriad ways.
\autoref{fig:design-space} summarizes the design space of such editors using several representative example systems, organized around two dimensions:
the primitive shapes used for layout, and whether or not these shapes can be nested horizontally and/or vertically.

\begin{figure}[b]
\newcommand{\STAB}[1]{\begin{tabular}{@{}c@{}}#1\end{tabular}}
  \begin{tabular}{cc|ccc}
     \multirow{5}{*}{\STAB{\rotatebox[origin=c]{90}{\textbf{Nestable?}}}}
     &&&&                                                                  \\
     & Horiz/Vert & Sandblocks          & Hass                  & \rocks{} \\
     & Horiz      & Fructure, tylr      & ---                   & ---      \\
     & No         & \emph{text editors} & \emph{text selection} & Deuce    \\
     &&&&                                                                  \\\hline
     &            & Rectangles \sba{}   & S-Blocks \sbc{}       & Rocks    \\
     & \multicolumn{1}{c|}{} & \multicolumn{3}{c}{\textbf{Primitive Shapes}}
  \end{tabular}
\caption{Design space of structure editors for code.}
\label{fig:design-space}
\end{figure}
 
The simplest shape for rendering nested elements is the rectangle.
In a typical flat code editor, such as CodeMirror, VS Code, or Vim, each line of text can be understood as a sequence of adjacent rectangles.
Compared to this line-based approach, in Sandblocks~\cite{Sandblocks2020, Sandblocks2023} rectangles can be arbitrarily nested; this approach completely reveals the text structure, at the cost of large layouts that differ significantly from the underlying, unstyled text.
In between such approaches are systems, such as Fructure~\cite{Fructure} and tylr~\cite{tylr}, which allow nested decorations to affect horizontal spacing but limit the use of vertical spacing---the fundamental challenge that stands in the way of compactly rendering nested elements.

Another type of primitive layout element, dubbed \emph{s-blocks} by \citet{Hass}, are rectangles but where the top-left and bottom-right corners are potentially missing.
As such, s-blocks resemble the way in which selected-text regions are displayed in many GUIs.
Their system, called Hass, includes a layout algorithm that generates s-blocks to render nested, potentially multiline strings.
Like ours, their algorithm attempts to minimally disturb the text layout as compared to unstructured text.
However, it is essentially a line-by-line approach, with judicious use of spacer glyphs throughout the text to give the appearance of nesting.
Their results are somewhat more compact than nested rectangles, but even on small examples the resulting layouts include undesirable visual artifacts, such as unusually large gaps between certain lines, and misalignment of characters from the preformatted text.

Compared to the four corners of a rectangle and at most eight corners of an s-block, arbitrary rectilinear polygons (which we call \emph{ragged blocks}) provide additional opportunities for tightly packing nested text. 
When hovering over the source text, the Deuce editor~\cite{Hempel2018} renders a ragged block that tightly wraps the corresponding substructure.
But unlike in \rocks{}, Deuce does not allow ragged blocks to be nested.

Beyond the axes in \autoref{fig:design-space}---shapes and ability to nest them---there are others we might use to categorize the diverse set of tools which visualize structured text.
The set of primitive glyphs, for example, can take many forms.
Projectional editors, such as MPS~\cite{MPS} and Hazel~\cite{OmarLivelits}, among others~\cite{OmarGraphite, Ko2006, Andersen2020, Erwig1995}, variously permit tables, images, and other domain-specific visuals to intermix with source text.
In this regard, the algorithms in this paper take a lower-level view of the world:
the leaf elements are rectangles, and clients of \rocks{} are responsible for drawing content within them, be it text or graphics.

Furthermore, the design space we present is concerned with the display a single ``node'' of text. ``Nodes-and-wires'' interfaces, such as Scratch~\cite{Resnick2009} and LabVIEW~\cite{LabVIEW}, among others, display multiple program fragments, which can be freely rearranged by the user on a two-dimensional canvas.
Manual positioning of multiple nodes is a separate concern from ours---the presentation of any single node of program text.

\paragraph{Interfaces for Program Analyzers}

Another rich source of examples are interfaces which are designed for the manipulation, analysis, or synthesis of programs and text.

The authors of LAPIS~\cite{LAPIS2002}, for example, explain how text editors could be extended to use multiple, potentially overlapping text selections.
They recognize a fundamental concern with rendering multiple selections using flat decorations, noting that two adjacent selections ``would be indistinguishable from a single selection spanning both regions.'' The STEPS~\cite{STEPS} system uses programming by example to extract structure from unstructured ``raw'' text.
Users can then perform queries and transformations on the structure-annotated output.
The system uses a blocks-like display for visualizing the extracted structure.

Inspired by LAPIS and STEPS, FlashProg~\cite{Mayer2015} is a tool for searching through the large space of specification-satisfying programs for structured text extraction.
The user interface developed for the system visualizes the structure of the extracted text with a line-by-line approach.
The authors also note the importance of nested text selections, lamenting that ``LAPIS unfortunately does not have good support for nested and overlapping regions, which are essential for data extraction tasks.''

Several other systems, such as reCode~\cite{reCode}, CoLadder~\cite{CoLadder}, and Ivie~\cite{Ivie}, use blocks, inline diffs, and other visualizations to help interactively generate and explain synthesized code fragments.

\paragraph{Prose Editors}

Some editors for natural language also visualize structure.
In Textlets~\cite{Textlets}, users can manipulate concurrent text selections to help maintain consistency, perform large edits, or generate summary information in legal documents.
In FFL~\cite{FFL}, users can apply flat text decorations and attach annotations to structural components of math formulas.
\rocks{} may help imbue these kinds of visualizations with even more structural detail.

\paragraph{Layout Algorithms}

A separate source of related works are those concerning layout in general, independent of particular application domains.
In \rocks{}, we adapt the TeX~\cite{BreakingParagraphs} line-breaking algorithm in our own algorithm (\Algos{L2b}) for rendering justified text.
The HTML and CSS box model~\cite{mdnBoxModel} supported by browsers provide the option to display rectangles either inline or as blocks, corresponding to the flat and nested extremes discussed above.
Penrose~\cite{Penrose} and Bluefish~\cite{Bluefish} are systems for specifying diagrams; the former uses numerical optimization to compute layouts, while the latter uses a more local constraint propagation model.

\section{Initial Layout Algorithm (\Algop{L1})}
\label{sec:version-1}

\begin{figure}[b]
  \includegraphics[width=3.3in]{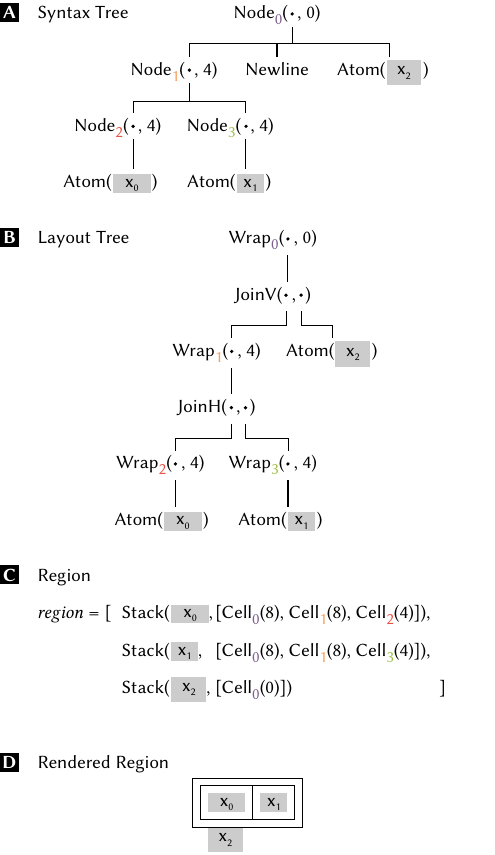}
  \vspace{0.15in} \caption{A syntax tree using \Unit{Node}s and \Unit{Newlines}, its corresponding layout tree using \Unit{Wrap}, \Unit{JoinV}, and \Unit{JoinH}, its derived region, and a rendering of the region.}
  \label{fig:pure-region-example}
\end{figure}

\newcommand{\sectionLabel}[1]
{\raisebox{-0.15em}{\textbf{\textit{#1}}}\hspace{0.04in}\hrulefill}

\newcommand{\subsectionLabel}[1]
{\begin{center}{\textbf{\textit{#1}}}\end{center}}

\newcommand{\termAtom}{\term{x}} \newcommand{\termNum}{\term{n}} \newcommand{\termId}{\term{id}}
\newcommand{\termPadding}{\term{padding}}
\newcommand{\termTree}{\term{t}} \newcommand{\termCells}{\term{cells}}
\newcommand{\termStack}{\term{stack}}
\newcommand{\termRegion}{\term{region}}
\newcommand{\termRegions}{\term{regions}}
\newcommand{\termLayout}{\term{layout}}
\newcommand{\termVector}{$\mathbf{v}$}
\newcommand{\termRegionVector}{\term{rv}}
\newcommand{\signatureSep}{\hspace{0.15in}}
\newcommand{\signatureOneArg}[3]{\fn{#1}~#2~=~#3}
\newcommand{\signatureTwoArgs}[4]{\fn{#1}~#2~#3~=~#4}
\newcommand{\signatureThreeArgs}[5]{\fn{#1}~#2~#3~#4~=~#5}

\newcommand{\doubleplus}{\ensuremath{+\kern-1.1ex+\kern0.3ex}}

\begin{figure*}

  \vspace{0.15in}

    \begin{center}
      \begin{tabular}{l c l}
        $t$ & \pdn{} &
        $\Constr{Atom}{x_i}$
        | $\Wrap{\term{id}}{t, \term{padding}}$
        | $\Constr{JoinV}{t, t}$
        | $\Constr{JoinH}{t, t}$ \\

$\term{stack}$ & \pdn{} & $\Stack{x_i, \termCells}$
         \hspace{0.0in} \text{where} \termCells~::= $\List{\Cell{\term{id}}{\term{$\Sigma_p$}}}$
        \\
        $\term{region}$ & \pdn{} & $\List{\term{stack}\hspace{0.01in}}$ \\

$\term{layout}$ & \pdn{} & $\List{\termRegionVector{}\hspace{0.01in}}$
         \hspace{0.0in} \text{where} \termRegionVector~::= (\termRegion, \termVector) \text{and} \termVector~::= $\langle\termNum, \termNum\rangle$
      \end{tabular}
    \end{center}
    \vspace{0.16in}

  \begin{minipage}{2in}
    \subsectionLabel{Advance \& Leading}
    \begin{flalign*}
      \signatureTwoArgs{spaceBetween}{\termCells}{\termCells}{& (\termNum, \termNum)} \\
      \signatureTwoArgs{advance}{\termStack}{\termStack}{& \termNum} \\[6pt]
      \signatureTwoArgs{leading$_\text{X}$}{\termAtom}{\termAtom}{& \termNum} \\
      \signatureTwoArgs{leading$_\text{S}$}{\termStack}{\termStack}{& \termNum} \\
      \signatureTwoArgs{leading$_\text{R}$}{\termRegion}{\termRegion}{& \termNum} \end{flalign*}
  \end{minipage}
\hspace{0.1in}
\begin{minipage}{2in}
    \subsectionLabel{Union \& Wrap}
    \begin{flalign*}
      \signatureTwoArgs{union$_\text{R}$}{\termRegion}{\termRegion}{& \termRegion} \\
      \signatureTwoArgs{union$_\text{L}$}{\termRegionVector}{\termRegionVector}{& \termRegionVector} \\[6pt]
      \signatureThreeArgs{wrap$_\text{S}$}{\termId}{\termPadding}{\termStack}{& \termStack} \\
      \signatureThreeArgs{wrap$_\text{R}$}{\termId}{\termPadding}{\termRegion}{& \termRegion} \\
      \signatureThreeArgs{wrap$_\text{L}$}{\termId}{\termPadding}{\termRegionVector}{& \termRegionVector} \end{flalign*}
  \end{minipage}
\hspace{0.1in}
\begin{minipage}{2in}
    \subsectionLabel{Layout}
    \begin{flalign*}
      \signatureTwoArgs{join$_\text{V}$}{\termRegion}{\termRegion}{& \termRegion} \\
      \signatureTwoArgs{join$_\text{H}$}{\termRegion}{\termRegion}{& \termRegion} \\[6pt]
      \signatureOneArg{layout}{\termTree}{& \termLayout} \\
      \signatureOneArg{merge}{\termLayout}{& \termRegion}
      \\
\end{flalign*}
  \end{minipage}

  \vspace{0.05in} 

  \caption{Syntax of Algorithm \Algop{L1}.}
  \label{fig:L1p-syntax}
\end{figure*}

\newcommand{\beforeNextSectionLine}{\vspace{6pt}}

\begin{figure*}
  \vspace{0.05in}

\sectionLabel{Advance}

{\begin{minipage}{0in} \begin{flalign*}
      \fn{spaceBetween}~[]~[]~&=~(0, 0) \\
\fn{spaceBetween}~[]~
      \Cell{\term{id}_b}{\Sigma_{pb}}:\term{rest}_b
      &= (0, \Sigma_{pb}) \\
\fn{spaceBetween}~
      \Cell{\term{id}_a}{\Sigma_{pa}}:\term{rest}_a~[]
      &= (\Sigma_{pa}, 0) \\
\fn{spaceBetween}~
      ( \Cell{\term{id}_a}{\Sigma_{pa}}:\term{rest}_a )~
      ( \Cell{\term{id}_b}{\Sigma_{pb}}:\term{rest}_b )
      &= \makebox[1.5in][l]{$(\Sigma_{pa}, \Sigma_{pb})$}
         \text{when } \term{id}_a \not= \term{id}_b \\
      &= \makebox[1.5in][l]{$\fn{spaceBetween}~\term{rest}_a~\term{rest}_b$}
         \text{otherwise} \\[8pt]
\fn{advance}~
      \Stack{x_a, \term{cells}_a}~
      \Stack{x_b, \term{cells}_b} &= \fn{width}~x_a + \Sigma_{pa} + \Sigma_{pb} \\
      &\qquad\quad\text{where }
      (\Sigma_{pa}, \Sigma_{pb}) = \fn{spaceBetween}~\term{cells}_a~\term{cells}_b \end{flalign*}
  \end{minipage}}
  \hfill

\beforeNextSectionLine
  \sectionLabel{Leading}

  {\begin{minipage}{0in} \begin{flalign*}
    \fn{leading}_\text{X}~x_i~x_j &=
       \makebox[1.2in][l]{$\fn{bottom}~x_i - \fn{top}~x_j$}
       \text{when } x_i \text{ and } x_j \text{ horizontally overlap} \\
&= \makebox[1.2in][l]{$0$} \text{otherwise} \\[8pt]
\fn{leading}_\text{S}~\Stack{x_i, \term{cells}_a}~\Stack{x_j, \term{cells}_b}
    &=\fn{leading}_\text{X}~(\fn{inflate}~x_i~\Sigma_{pa})~(\fn{inflate}~x_j~\Sigma_{pb}) \\
    &\qquad\quad\text{where } (\Sigma_{pa}, \Sigma_{pb}) = \fn{spaceBetween}~\term{cells}_a~\term{cells}_b \\[8pt]
\fn{leading}_\text{R}~\term{region}_a~\term{region}_b
    &=\fn{maximum}~[\fn{leading}_\text{S}~\term{stack}_i~\term{stack}_j~|~ \text{for each } \term{stack}_i\in\term{region}_a~\text{and}~\term{stack}_j\in\term{region}_b]
  \end{flalign*}
  \end{minipage}}

\beforeNextSectionLine
  \sectionLabel{Union \& Wrap}

  {\begin{minipage}{0in} \begin{flalign*}
    \unionr{}~a~b &= a~\doubleplus{}~b \\ \unionl{}~(a, \mathbf{v}_a)~(b, \mathbf{v}_b) &= (\unionr{}~a~b, \mathbf{v}_a + \mathbf{v}_b) \\[8pt]
    \wraps{}~\term{id}~\term{padding}~\Stack{x_i, \term{cells}}
    &= \Stack{x_i, \Cell{\term{id}}{\term{padding}}:\term{cells}} \\\wrapr{}~\term{id}~\term{padding}~\termRegion
    &=
    [\wraps{}~\term{id}~\term{padding}~\term{stack}\hspace{0.01in}~|~\text{for each \term{stack} in \termRegion}\hspace{0.01in}] \\
    \wrapl{}~\term{id}~\term{padding}~(region, \mathbf{v})
    &=
    (\term{region}',~\mathbf{v} + \langle2 \times \term{padding}, 0\rangle) \\
    &\qquad\quad\text{where $\term{region}'=(\wrapr{}~\term{id}~\term{padding}~\term{region})$ translated right by \term{padding}}
  \end{flalign*}
  \end{minipage}}

\beforeNextSectionLine
  \sectionLabel{Layout}

  {\begin{minipage}{0in} \begin{flalign*}
      \joinv{}~a~b &= a~\doubleplus{}~b \\ \joinh~a~b &=
      \makebox[1.12in][l]{$a~\doubleplus{}~b$}
      \text{when } a = [] \text{ or } b = [] \\
      \joinh~a~b &=
      \makebox[1.12in][l]{$a'~\doubleplus{}~[\termRegionVector\hspace{0.01in}]~\doubleplus{}~b'$}
      \text{otherwise } \\ &\qquad\text{where} \\
      &\qquad\qquad\text{$a'$ (resp. $b'$) are all but the last (resp. first) line of $a$ (resp. $b$),} \\
&\qquad\qquad\text{$\termRegionVector$ is the \unionl{} of the last line of $a$ and $b_\text{T}$, and} \\
      &\qquad\qquad\text{$b_\text{T}$ is the first line of $b$ translated by the last line of $a$'s advance} \\[8pt]
\fn{layout}~\Constr{Atom}{x_i} &= \List{(\List{\Constr{Stack}{x_i, \List{}}}, \langle \fn{width}~x_i, 0\rangle)} \\
    \fn{layout}~\Constr{JoinV}{t_1, t_2} &= \fn{join}_\text{V}~(\fn{layout}~t_1)~(\fn{layout}~t_2) \\
    \fn{layout}~\Constr{JoinH}{t_1, t_2} &= \fn{join}_\text{H}~(\fn{layout}~t_1)~(\fn{layout}~t_2) \\
\fn{layout}~\Wrap{\term{id}}{t, \term{padding}} &= [\fn{wrap}_\text{L}~\term{id}~\term{padding}~\termRegionVector\hspace{0.01in}~|~\text{for each \termRegionVector{} in (\fn{layout}~t)}]
    \\[8pt]
    \fn{merge}~\termLayout &= \fn{merge\_}~[\termRegion\hspace{0.01in}~|~\textrm{for each (\termRegion, \termVector) in \termLayout}\hspace{0.01in}]~[] \\
\fn{merge\_}~\List{}~\term{done} &= \term{done} \\
    \fn{merge\_}~(\termRegion:\termRegions)~\term{done} &= \fn{merge\_}~\termRegions~(\fn{union}_\text{R}~\term{done}~\termRegion{}') \\
    &\qquad\text{where $\termRegion{}' = \termRegion$ translated down by $(\fn{leading}_\text{R}~\term{done}~\termRegion)$}
\end{flalign*}
    \end{minipage}}

\vspace{0.05in} 

\caption{Algorithm \Algop{L1}.}
  \label{fig:L1p}
\end{figure*}

In order to describe text layout, it is convenient to forget that the fundamental objects to be laid out are words or letters, and instead consider only rectangles.
The layout algorithm need not be concerned with what is \emph{inside} these rectangles.
There could be a single letter, a whole word, a sequence of words, or even an image or an interactive visualization.
Importantly, these rectangles will be considered \emph{indivisible} for the purposes of layout.
To differentiate these atomic rectangles from, say, the rectangles in a boxes layout, we will refer to them as \emph{fragments}.

Let us forget for a moment that we care about rendering structured text, and consider the case of \emph{unstructured} text.
Then, we could represent a text layout by the following grammar:

\begin{center}
\begin{tabular}{l c l}
$u$ & \pdn{} & \List{\term{atom}} \\
$\term{atom}$ & \pdn{} & $x_i$ | \Unit{Newline}
\end{tabular}
\end{center}

An unstructured layout, $u$, is a sequence of \term{atom}s, each of which is either a fragment, $x_i$, or a \Unit{Newline}.

Then, finding the position of each $x_i$ amounts to laying out each $x_i$ in sequence, placing the left edge of each $x_{i+1}$ at position $\fn{left}~x_i + \fn{width}~x_i$, unless we encounter a \Unit{Newline}, in which case the left edge of the next $x_i$ is 0.
The amount that we translate a fragment $x_{i+1}$ from the left side of its left neighbor is called the \emph{advance} of the fragment.
In the case of unstructured text, $\fn{advance}~x_i~=~\fn{width}~x_i$, but we will retain the notion of $\fn{advance}$ since it more easily generalizes to the case of structured text layout.

We can describe the input to a structured text layout algorithm by enriching the input grammar with a constructor for \Unit{Node}s:

\begin{center}
\begin{tabular}{l c l}
$e$ & \pdn{} & $\Constr{Atom}{\term{atom}}$ | $\Node{\term{id}}{\List{\term{e}}, \term{padding}}$ \\
$\term{atom}$ & \pdn{} & $x_i$ | \Unit{Newline}
\end{tabular}
\end{center}

\noindent
Here, $e$ is a \emph{syntax tree}, and $x_i$ are fragments.
\autoref{fig:pure-region-example}~\letter{a} shows an example syntax tree for a small text layout involving three fragments.
The grammar of syntax trees is useful for expressing a tree structure inherent to a document, but it is not convenient for specifying the semantics of structured text layout.
The problem is that the constructors for \Unit{Node} and \Unit{Newline} are not on equal footing; \Unit{Node} is an arbitrary node in the tree (potentially containing children), whereas \Unit{Newline} is a leaf node (serving as a delimiter in a list of subtrees).
In a rocks layout, however, \Unit{Node}s and \Unit{Newline}s are equivalently important sources of structure.
It is convenient, then, to treat \Unit{Node}s \emph{and} \Unit{Newline}s as non-leaf nodes.

The grammar of \emph{layout tree}s, $t$, given in \autoref{fig:L1p-syntax} does just this, introducing the \Unit{JoinV}, \Unit{JoinH}, and \Unit{Wrap} constructors, which together supplant the combination of \Unit{Node} and \Unit{Newline}.
\autoref{fig:pure-region-example}~\letter{b} shows the layout tree which corresponds to the syntax tree given in \autoref{fig:pure-region-example}~\letter{a}.
This grammar breaks \Unit{Node}s into two distinct concepts: the concept of \Unit{Wrap}-ping a subtree in some padding, and the horizontal concatenation of subtrees (denoted by \Unit{JoinH}).
\Unit{Newline}s are expressed by vertical concatenation (denoted by \Unit{JoinV}).

Translation from syntax to layout tree is a straightforward re-parsing of the children in each \Unit{Node}.
We can obtain an equivalent layout tree by treating the children of each \Unit{Node} in the syntax tree as a sequence of tokens, and parsing the token list as an application of infix operators, \Unit{JoinH} between two non-newline nodes, and \Unit{JoinV} between a node and a \Unit{Newline}.

Given a layout tree, the purpose of layout is to determine the location of its leaves (\ie{}~the fragments), and furthermore the dimensions of a ragged block surrounding each \Unit{Node}.
Our algorithm involves several operations on a data structure, called regions, and several options on layout trees; we describe each group below.

\subsection{Regions}

We can define the semantics of a layout over a layout tree by way of analogy to layout over the unstructured text.
We do this first by re-defining advance as an operation which considers the local padding around a fragment.

\paragraph{Advance}

Unfortunately, we cannot define advance in terms of the raw fragments. We need to construct an auxiliary data structure containing information about the ancestry (in the layout tree) of each fragment.
This is because the space between two fragments in the layout is a function of the fragments' position in the layout tree.
Two sibling fragments (like $x_0$ and $x_1$ in \autoref{fig:pure-region-example}) which are wrapped by a common ancestor ($\Unit{Wrap}_1$ in \autoref{fig:pure-region-example}), ought to be placed closer together than two fragments which don't share the ancestor (such as $x_0$ and $x_2$, for example).
This is the mechanism by which we can allocate space for borders around \Unit{Wrap}-ped nodes in the layout.
We call this auxiliary data structure a \term{region}.

A region is a list of \Unit{Stack}s (one for each text fragment), and a \Unit{Stack} is a fragment paired with a list of \Unit{Cell}s (the list representing a path in the layout tree).
Each \Unit{Cell} records a \Unit{Wrap} node which occurs above its \Unit{Stack}'s fragment in the layout tree.
For example, in \autoref{fig:pure-region-example}, the \Unit{Stack}s for fragments $x_0$ and $x_1$ each have three \Unit{Cell}s since a path from $x_0$ or $x_1$ up the layout tree to the root encounters three \Unit{Wrap} nodes.
The argument to a \Unit{Cell} records the \emph{cumulative} padding applied due to a \Unit{Wrap} node (see $\Unit{Cell}_0$, for example, in \autoref{fig:pure-region-example}; its corresponding \Unit{Wrap} node applies \emph{no} padding, and so $\Unit{Cell}_0$'s argument is \emph{unchanged} from its successor).
By storing the cumulative padding, the wrapped rectangle corresponding to a \Unit{Stack} can be found without chasing the list of \Unit{Cell}s.

We provide here two plausible (but incorrect) definitions of advance which illustrate the importance of the \Unit{Cell} in determining if two \Unit{Stack}s are \emph{compatible}.

\begin{center}
\begin{minipage}[b]{2in}
\flushleft
$\fn{advance}_\text{unsound}$\\
\qquad\Constr{Stack}{$x_a$, \Constr{Cell}{\term{$\term{id}_a$}, $\Sigma_{pa}$}:$\term{rest}_a$}\\
\qquad\Constr{Stack}{$x_b$, \Constr{Cell}{\term{$\term{id}_b$}, $\Sigma_{pb}$}:$\term{rest}_b$}\\
\qquad\qquad = \fn{width}($x_a$)
\end{minipage}
\includegraphics[width=1in]{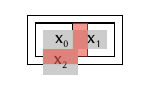}

\begin{minipage}[b]{2in}
\flushleft
$\fn{advance}_\text{conservative}$\\
\qquad\Constr{Stack}{$x_a$, \Constr{Cell}{\term{$\term{id}_a$}, $\Sigma_{pa}$}:$\term{rest}_a$}\\
\qquad\Constr{Stack}{$x_b$, \Constr{Cell}{\term{$\term{id}_b$}, $\Sigma_{pb}$}:$\term{rest}_b$}\\
\qquad\qquad = \fn{width}($x_a$) + $\Sigma_{pa}$ + $\Sigma_{pb}$
\end{minipage}
\includegraphics[width=1in]{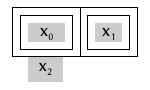}
\end{center}

The first definition is a naive re-interpretation of our notion of advance from flat text layout which ignores padding completely.
This leads to unsound layouts because space allocated for a \Unit{Wrap} in the layout tree might overlap unrelated rectangles.
The second definition takes the outermost, maximal padding at each \Unit{Cell}, and says that the \fn{advance} is the width of $x_a$ plus the maximal padding that can occur between $x_a$ and $x_b$.
This definition of \fn{advance} is sound, but it is too conservative.
Elements in the layout tree which are underneath the same \Unit{Wrap} node ought to share space.

The proper definition of advance in \autoref{fig:L1p} uses a helper function, \fn{spaceBetween}, which traverses the \Unit{Stack}s in parallel, shedding \Unit{Cell}s which derive from the same \Unit{Wrap} node.
Alternatively, we can think of \fn{spaceBetween} as finding the padding due to the lowest common ancestor \Unit{Wrap} node of the arguments.\footnote{Strictly speaking, the padding due to the lowest common ancestor (LCA) \Unit{Wrap} node, \emph{plus} the padding ascribed to any \Unit{Wrap} nodes on a path from the LCA to the root of the tree, since a \Unit{Cell} stores \emph{cumulative} padding, not the padding due to a single \Unit{Wrap}.
}
Once we reach the end of the stack, or a pair of \Unit{Cell}s which derive from different parts of the layout tree (as determined by the \Unit{Cell}'s \term{id}, which corresponds to the \term{id} of a \Unit{Wrap} node in the layout tree), the cumulative padding of these \Unit{Cell}s is the padding that must be applied between the fragments.

\paragraph{Leading}

Careful readers of the definition of \fn{spaceBetween} may note that this operation doesn't consider the dimension of the rectangles associated with each cell.
The minimum space between two fragments is \emph{only} a function of the fragment's \Unit{Stack} (and, hence, the \emph{structure} of the layout tree), not the dimension or position of the fragment.
Therefore, we can use \fn{spaceBetween} to not only decide the \emph{horizontal} spacing of rectangles (\ie{}~\fn{advance}), but also the \emph{vertical} space between rectangles, known as the \emph{leading}.\footnote{We use the term leading in this paper to mean the space between baselines (as opposed to the inter-line spacing).}

We define $\fn{leading}_\text{X}~x_i~x_j$ as the amount of space needed to put rectangle $x_j$ entirely below $x_i$, or 0 if $x_i$ and $x_j$ don't overlap horizontally.
This definition easily extends to $\Unit{Stack}$s by finding two rectangles which are centered on $x_i$ and $x_j$, but are ``inflated'' (translating all edges outwards by the same amount) by the space between $x_i$ and $x_j$ respectively.
The leading between two regions is the maximum of the $\fn{leading}_\text{S}$ between each pair of stacks in the regions.

The process of layout will combine regions together until we have a single region representing each line.
Then, these lines are combined into a single region using $\fn{merge}$.
The $\fn{leading}_\text{R}$ function is used to find the minimum space to leave between two successive lines such that no two \Unit{Stack}s improperly overlap.

\paragraph{Union \& Wrap}

But how do we actually \emph{combine} regions?
One of the principal advantages of representing regions as an implicit union of rectangles is that combining regions is equivalent to concatenating their \Unit{Stack}s.
The order of a region's stacks does not matter (but as we'll see in \autoref{sec:simplification} it can be exploited), and we make no attempt to resolve collisions between stacks (this is the concern of layout, which we will address next).

\begin{figure}
\includegraphics[width=3.3in]{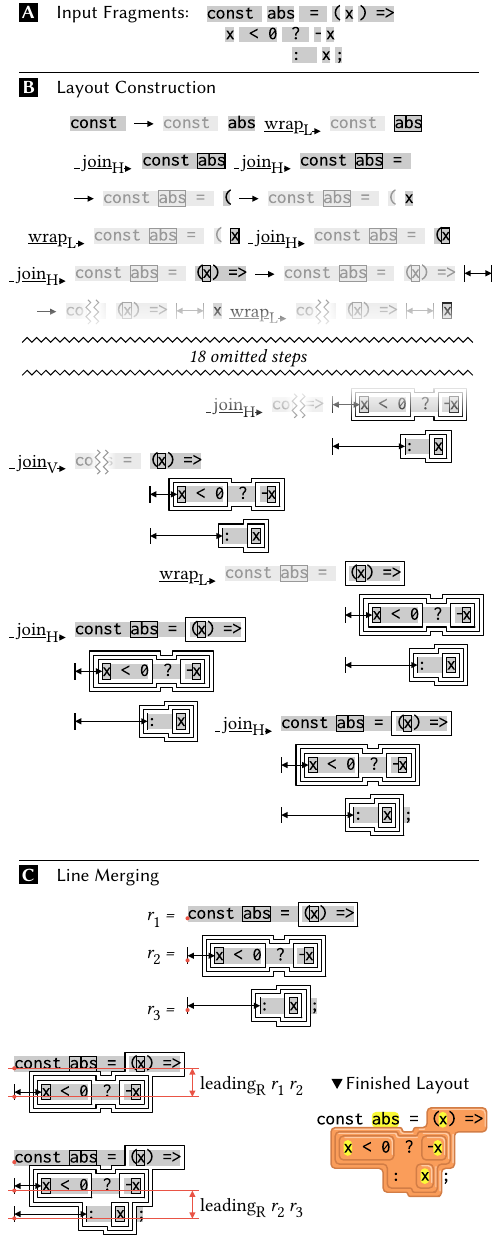}
\caption{Example \Algop{L1} layout steps.}
\label{fig:L1p-reduction}
\end{figure}
 
\newcommand{\origin}{\raisebox{0in}{\includegraphics{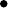}}}
\newcommand{\preTransOrigin}{\raisebox{0in}{\includegraphics{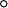}}}

\begin{figure}[t]
\includegraphics[width=3.3in]{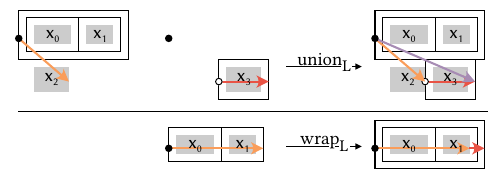}
\caption{Example applications of $\fn{union}_\text{L}$ and $\fn{wrap}_\text{L}$.
The world origin is depicted by \origin{}, while the pre-translation origin is depicted by \preTransOrigin{} (shown only for illustrative purposes).
}
\label{fig:wrap-and-union}
\end{figure}

\subsection{Layout}

Algorithm \Algop{L1} works analogously to layout of unstructured, pre-formatted text.
First, each line is laid out individually through the repeated application of horizontal and vertical concatenation and wrapping operations.
\autoref{fig:L1p-reduction} shows the application of these operators to produce a layout for \Abs{}.
The lines, once laid out, are combined at the end to form a finished layout (see \autoref{fig:L1p-reduction} \letter{c}).

The type of an \Algop{L1} layout is defined in \autoref{fig:L1p-syntax}.
It is a list of lines, each line represented by a single region.
Alongside each region is a vector, $\mathbf{v}$, which is the advance of the region.
The advance vector records the position, relative to a region's origin, that a region following it should be placed.
When two regions in the layout are combined with \unionr{}, or a region in the layout is wrapped, the advance vector is updated accordingly.
\autoref{fig:wrap-and-union} shows an example of how the advance vector is updated for each of these operations.

As defined in \autoref{fig:L1p}, layout proceeds as a structural recursion over the layout tree, \term{t}.
During this recursion, we apply one of two join operations, \joinh{} and \joinv{}.
These join functions combine two layouts into a single layout, either joining them horizontally (\joinh{}) or vertically (\joinv{}).
Vertical concatenation amounts to concatenation of the layout's lines.
Horizontal concatenation splices the two abutting lines together, using \fn{union} to combine them.

\autoref{fig:L1p-reduction} gives several examples of \joinh{} and \joinv{} in action.
The figure can be read as a sequence of stack operations, representing a possible order of operations in the recursive evaluation of layout.
Each element of the stack is a single layout, and the top of the stack is on the right.
\joinh{} and \joinv{} operate on the two topmost layouts in the stack, joining them either horizontally or vertically respectively.
(In the case where a new element is pushed and then immediately joined with \joinh{}, the push is omitted.)

\subsection{Spacers}

\begin{figure}[t]
\includegraphics[width=1.5in]{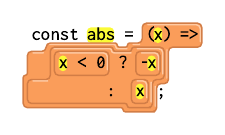}
\caption{Example \Algop{L1} layout without spacers.}
  \label{fig:L1p-result-no-spacers}
\end{figure}

The algorithm given in \autoref{fig:L1p} is enough to produce usable structured text layouts, but it does not elegantly handle whitespace.
Specifically, preceding whitespace on a line \emph{must} be represented by a fragment.
Furthermore, this fragment is not exempt from wrapping.
In practice, this results in unsightly layouts where preceding whitespace draws unneeded visual attention (\eg{}~see \autoref{fig:L1p-result-no-spacers}).

To fix this visual defect, we introduce a new kind of leaf called a \Unit{Spacer}, which
behaves similarly to an \Unit{Atom} except it has no height and does not participate in wrapping.
Its width is given by its scalar argument, $w_i$, which is normally set to the combined width of the whitespace characters it represents.
\autoref{fig:L1p-spacers} details the modifications to \Algop{L1} needed to support this new object.
The \Unit{Spacer}'s invariance to wrapping is implemented in the new definition of \fn{advance}, which ignores the \fn{spaceBetween} any object and a \Unit{Spacer}.

\newcommand{\afterNextSectionLine}{\vspace{-5pt}}

\begin{figure}[t]
  \vspace{0.10in}

    \begin{center}
      \begin{tabular}{l c l}
$t$ & \pdn{} &
$\cdots$
        | \new{$\Constr{Spacer}{w_i}$} \\

$\term{stack}$ & \pdn{} & $\cdots$
        | \new{$\Constr{Spacer}{w_i}$} \\
      \end{tabular}
    \end{center}

\sectionLabel{Advance}
    \afterNextSectionLine

    \begin{flalign*}
&\new{$\fn{advance}~
      \Stack{x_a, \term{cells}_a}~
      \Constr{Spacer}{w_b} = \fn{width}~x_a$} \\
      &\hspace{0.76in}\new{$\fn{advance}~
      \Constr{Spacer}{w_a}~y = w_a$}
    \end{flalign*}

\sectionLabel{Leading}
    \afterNextSectionLine

    \begin{align*}
&\new{$\fn{leading}_\text{S}~x~y = 0$}
    \end{align*}

\sectionLabel{Layout}
    \afterNextSectionLine

    \begin{flalign*}
      &\new{$\fn{layout}~\Constr{Spacer}{w_i} = \List{(\List{\Constr{Spacer}{w_i}}, \langle w_i, 0\rangle)}$}
    \end{flalign*}

\caption{Algorithm \Algop{L1}, continued (with spacers).}
  \label{fig:L1p-spacers}
\end{figure}

\section{Extensions}

Next, we describe several extensions to the basic algorithm.
We sketch the main ideas with less detail than in the previous section.

\subsection{Layout with Column Constraints (\Algop{L2a})}
\label{sec:extensions-2a}

The finished layout produced by \Algop{L1} (see \autoref{fig:L1p-reduction}) succeeds in visualizing the structure of the document.
When writing code, authors often align certain columns of text to aid readability, or draw attention to some symmetry between adjacent lines of source text.
In the case of \autoref{fig:L1p-reduction}, the original source program aligned the ``\texttt{?}'' and ``\texttt{:}'' characters, but this alignment was destroyed by the addition of padding.
Layouts could be improved further by maintaining additional constraints regarding \emph{formatting} in code.

The \rocks{} layout shown in \autoref{fig:blocks-vs-rocks}, by contrast, was generated by algorithm \Algop{L2a}, which maintains the alignment of columns through the use of constraints.
In algorithm \Algop{L2a}, the \fn{spaceBetween} two fragments is not used to directly set the position of the fragments, but to generate a \emph{lower bound constraint}, ensuring that no fragments come \emph{closer than} the \fn{spaceBetween} them, but are permitted to space apart if necessary.
The layout tree is modified to include a notion of named constraint variables, allowing the client to specify constraints on the horizontal position of fragments in the layout.
\refAppendixVersionTwoA{} discusses this extension in more detail.

\begin{figure}[t]
\includegraphics[width=3.3in]{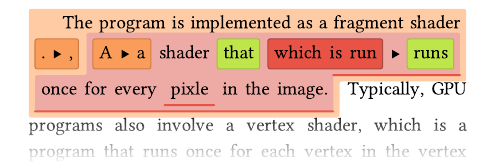}
  \caption{A hypothetical word processor visualizing three nested errors using rocks rendered by \Algop{L2b}: a spelling error (underlined), a grammatical error (red and green), and a suggestion to merge two sentences into one (orange).}
  \label{fig:L2bp-nested-errors}
\end{figure}
 
\subsection{Layout with Automatic Line Breaks (\Algop{L2b})}
\label{sec:extensions-2b}

Algorithms \Algop{L1} and \Algop{L2a} take as input {formatted} text, where the position of newlines is known.
But, it is common that when laying out prose, the document author does not choose explicit newlines.
Algorithm \Algop{L2b} includes a pre-processing step which finds good locations for line breaks, inserts them, generating a new layout tree, then generates a layout of the resulting tree.

Algorithm \Algop{L2b} is based on the algorithm by \citet{BreakingParagraphs} for choosing line breaks in a paragraph.
Because this algorithm, too, defines a notion of advance between two adjacent fragments, it is straightforward to substitute our new notion of advance, which considers the space between two fragments from their relationship in the layout tree.
\autoref{fig:L2bp-nested-errors} demonstrates an example of \Algop{L2b} in action.
Note that in order to justify the text, the space between fragments on each line have been shortened or lengthened as necessary.
A more complete discussion of \Algop{L2b} can be found in \refAppendixVersionTwoB{}.

\subsection{Stateful Regions (\Algos{L1}, \Algos{L2a}, \Algos{L2b})}
\label{sec:stateful-regions}

The pure regions presented in \autoref{sec:version-1} are conceptually simple and convenient, but they are not efficient in practice.
This is especially true if we need to maintain several regions simultaneously.
The algorithms presented so far have no cause to do this, but as we'll see in \autoref{sec:simplification}, it is convenient to be able to annotate a node in the layout tree with the region it corresponds to.
However, pure regions make this costly due to a lack of sharing.

\begin{figure}[b]
  \includegraphics[width=3.3in]{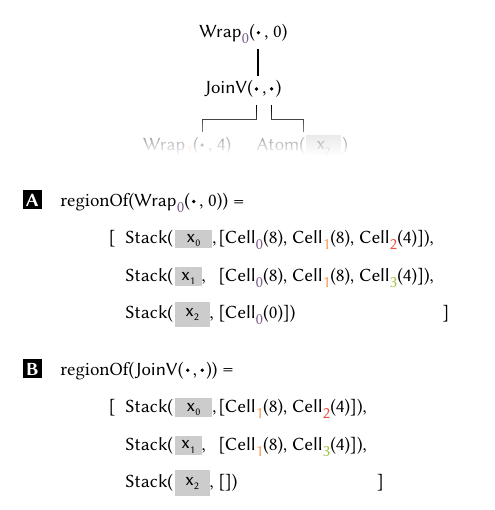}
  \caption{Regions of \autoref{fig:pure-region-example}~\letter{b}, root and child.}
  \label{fig:region-of}
\end{figure}

For example, \autoref{fig:region-of} shows the regions due to the root and the root's child in the small example given in \autoref{fig:pure-region-example}.
Most of \autoref{fig:region-of} \letter{a} and \letter{b} are the same; the difference being that the fragments under the former region lack \Unit{Cell}s due to $\Unit{Wrap}_0$.
In fact, it is always the case that the \Unit{Stack}s in a parent node $t$ will contain all of the \Unit{Cell}s of their children.
Stateful regions capitalize on this fact in order to reduce duplication between regions.

The generous sharing between pure regions in a layout also has a consequence on a \fn{translate} operation, not shown in \autoref{fig:L1p}.
If we \emph{do} annotate each node in the layout tree with its region, \fn{translate} becomes costly, the location of each fragment, $x_i$ being duplicated among nodes.

Stateful regions---which we swap in for pure regions, resulting in the algorithms \Algos{L1}, \Algos{L2a}, \Algos{L2b}---solve these problems by taking advantage of two observations about how regions are used in layout.
The first observation is that during layout, no rectangles (and by extension, no \Unit{Stack}s) are added or removed.
Said another way, it is possible, by traversing the layout tree prior to layout, to know the complete set of atomic fragments which will be involved in layout.
We can use this fact to modify the representation of \Unit{Stack}s so that they contain a \emph{reference} to the rectangle they wrap, rather than a representation of the rectangle itself:

\begin{center}
  \begin{tabular}{l c l}
    $\term{region}$ & \pdn{} & $\List{\term{stack}}$\vphantom{\changed{[}} \\
    $\term{stack}$ & \pdn{} & $\Constr{Stack}{\changed{$i$}, \List{\Constr{Cell}{\term{id}, \term{$\Sigma_p$}}}}$ \\
\new{$\term{rects}$} & \pdn{} & \new{$\List{x_0, \dots, x_n}$} \\
  \end{tabular}
\end{center}

\noindent
Now, each \Unit{Stack} points into a rectangle vector, \term{rects}, which is shared between every region.
This change obviates the need to traverse the layout tree during layout translations, since a translation of a layout node has the side effect of translating all of the regions in any layout below it.

The second observation is that layout never takes the union of two regions which are not adjacent.
Indeed, if layout \emph{did} take the union of two non-adjacent regions, it would correspond to a re-ordering, omission, or duplication of fragments in the underlying text, which are all operations which should obviously be avoided.
We can capitalize on this observation by representing regions not as a list of \Unit{Stack}s, but as a span in a table of \Unit{Cell}s which we call the \emph{timetable}. (So named because each column of the table is the \emph{schedule} of wrap operations which will occur on each fragment over the course of layout.)
To maintain the partial persistence of the region data structure, invoking $\fn{wrap}_R$ on a region mutates the table by adding a new row, thus ensuring that regions referring to less-wrapped spans are unaffected.\footnote{We call the stateful region \emph{partially} persistent because the effects of translations are visible to existing regions, but the effects of \fn{wrap}s are not.
}
\begin{center}
  \begin{tabular}{l c l}
    $\term{region}$ & \pdn{} & $\changed{\Constr{Region}{\term{begin}, \term{end}, \term{depth}}}$ \\
  \end{tabular}
\end{center}

\begin{figure}[b]
  \includegraphics[width=3.3in]{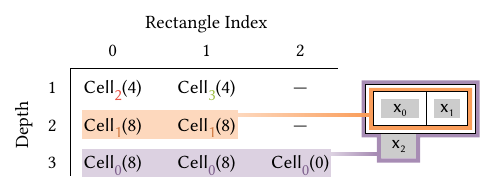}
  \caption{Timetable for \autoref{fig:pure-region-example} example.}
  \label{fig:timetable}
\end{figure}

\autoref{fig:timetable} shows the timetable that would be generated for the small layout given in \autoref{fig:pure-region-example}.
(A larger example is given in \refAppendixTimetableAbs{}.)
Two example regions corresponding to $\Unit{Wrap}_0$ (purple) and $\Unit{Wrap}_1$ (orange) are highlighted in the rendered output along with their span in the timetable.

The timetable is a function only of the structure of the layout tree; it does not depend on the size or position of the layout's rectangles.
This important quality means that the timetable can be constructed \emph{before} layout, by traversing the layout tree and ``simulating'' the effect of wrapping and vertical and horizontal concatenation.

\begin{figure*}[t]
  \includegraphics[width=7in]{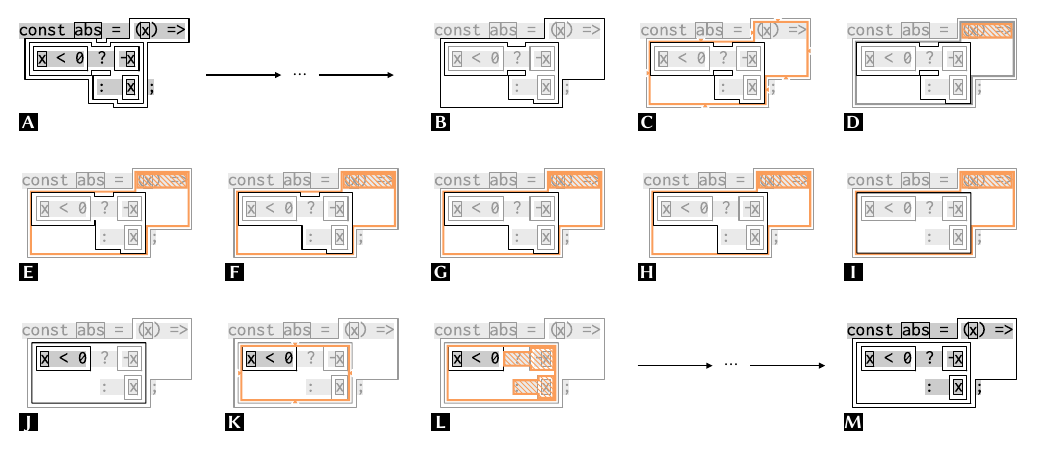}
  \caption{Excerpted steps of the simplification of \Abs{}.}
  \label{fig:simplification}
\end{figure*}

The timetable bears strong resemblance to the pure region formulation; each column contains the same \Unit{Cell}s that a \Unit{Stack} corresponding to the same fragment would contain.
However, whereas \Unit{Stack}s are permitted to have different lengths, each column in the timetable has the same depth.
Furthermore, some \Unit{Cell}s in the timetable are duplicated within a single column.
This is to ensure that every region corresponds to a span in a \emph{single row} in the timetable. (For example, the outermost region in \autoref{fig:timetable}, corresponding to $\Unit{Wrap}_0$ occupies a single row, despite the fact that its constituent rectangles have been wrapped between 1 and 3 times.)
The single-row property is enforced by a simple rule: when building the timetable, whenever a vertical or horizontal concatenation is encountered, the columns corresponding to the regions on the left-hand side and right-hand side of the operator are ``filled'' to the same depth by repeating the topmost element (or an empty element, if the column is empty, rendered as ``---'' in \autoref{fig:timetable}).
  
\section{Simplification}
\label{sec:simplification}

The layouts we've demonstrated so far succeed in visually presenting the structure of the underlying text, but the polygonal outlines of the regions generated by layout can be rather intricate. According to usage scenarios and user preferences, we might like to generate outlines which continue to visualize structure, but with less ragged edges (\ie{}~fewer corners) if possible. Simplification is a way to approach this goal.

However, when considering the simplicity of a rock, we are \emph{not} concerned with the simplicity of the underlying region representation (the union of a set of small rectangles), but rather the simplicity of the \emph{boundary polygon} which is a result of taking the union of this set.
Furthermore, the objects that we are concerned with in simplification---corners, intersections, edges, and so on---are not the objects that are easy to describe using regions.
For this reason, simplification will be an operation on rectilinear polygons, not regions.
It is a post-processing step that occurs only after layout.

An additional concern arises when we introduce rectilinear polygons as another object in the pipeline of rocks layout and presentation.
In particular, it will be necessary to phrase operations in terms of both the region being simplified (which is not always the topmost, finished region, but perhaps the region due to some subtree in \term{t}), and the layout tree.
For example, we might want to say, ``simplify a region, but ensure that its bounds do not exceed the bounds of its parent.''
This is not possible to express using only one region (we have a region to be simplified, and a region to stay inside of), and it is not possible to express if we forget the structure of the tree (since we mention a \emph{parent}, which can only refer to a node in the layout tree).

As mentioned in \autoref{sec:stateful-regions}, pure regions are costly to associate with nodes in the layout tree.
So, a practical implementation of simplification is predicated on the use of stateful regions, so that we might efficiently perform the required annotation.

\subsection{An Example}
\autoref{fig:simplification} gives an example of simplifying \Abs{}, as rendered by algorithm \Algos{L2b}.
Simplification is a top-down process which is implemented as a tree traversal over the layout tree.

At every stage in the simplification process, two polygons are maintained: a \term{keepIn} polygon, which describes the boundary that the region should say inside of, and a \term{keepOut} polygon, which describes an area which is ``off limits'' for the region being simplified.
In \autoref{fig:simplification}, the \term{keepIn} polygon is rendered with a thick orange border, and the \term{keepOut} polygon is rendered as a cross-hatched area.
\autoref{fig:simplification}~\letter{a} shows the unsimplified output of layout.
\autoref{fig:simplification}~\letter{b} shows a step of simplification just after the outermost outline has been fully simplified.
This outermost outline serves as the \term{keepIn} polygon for simplification of its child subtree.
\autoref{fig:simplification}~\letter{c} shows the new \term{keepIn} polygon, which has been offset by the padding applied to the outermost outline in the layout.
This ensures that even after simplification, a parent node's border is unoccupied by the outlines of its children.
\autoref{fig:simplification}~\letter{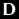} shows the \term{keepOut} polygon for this stage of simplification.
The \term{keepOut} polygon is derived from the \emph{sibling} elements of the term currently being simplified.

Once the \term{keepIn} and \term{keepOut} polygons are resolved, counter-clockwise corners and antiknobs (\ie{}~sides connecting two concave corners~\cite{BarYehuda}) are removed from the current rock.
\autoref{fig:simplification}~\letter{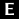} through \letter{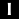} show the removal of these features.
The finished outline serves as the \term{keepIn} polygon of the subtree's simplification, as shown in \autoref{fig:simplification}~\letter{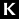}.
The final outlines after the recursion has been exhausted are shown in \autoref{fig:simplification}~\letter{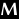}.

\subsection{Core Operations}
\label{sec:core-operations}

The simplification process is a top-down traversal of the layout tree, but also requires access to the region associated with each node (so that we might construct a polygonal outline from it).
In order to facilitate the retrieval of the relevant region, we can annotate each node in the layout tree with its corresponding region.
\begin{center}
  \begin{tabular}{l c l}
    $t$ & \pdn{} &
    $\Constr{Atom}{x_i}$
    | $\Wrap{\term{id}}{\changed{$T$}, \term{padding}}$ | \\ & &$\Constr{JoinV}{\changed{$T$}, \changed{$T$}}$
    | $\Constr{JoinH}{\changed{$T$}, \changed{$T$}}$ | $\Constr{Spacer}{w_i}$ \\[2pt]

    \new{$T$} & \pdn{} & \new{$(t, \term{region})$} \end{tabular}
\end{center}
Generally speaking, our goal is to modify the rectilinear polygonal outlines of each region in the layout tree by removing both anti-knobs and counter-clockwise wound corners (we assume that the outlines are wound clockwise).
However, we cannot perform these modifications without restriction.
In order to maintain that the tree structure that the layout presents \emph{before} simplification is the same as the structure it presents \emph{after} simplification, we enforce the following two rules:
\begin{enumerate}
  \item The outline of a child node shall not exceed the outline
    of its parent, less the padding applied to its parent.
  \item For a given node of the tree, the outlines of the node's
    children may not intersect.
\end{enumerate}
The first rule ensures that simplification does not permit a child node's boundary to cross the boundary of its parent.
In fact, it is slightly more restrictive; a child boundary is restricted to a boundary \emph{smaller than} that of its parent, to account for the padding applied to the parent node.

Whereas the first rule is concerned with the parent-child relationship between nodes, the second rule is concerned with the relationship between siblings.
If two siblings do not intersect prior to simplification, we ensure that they continue to be disjoint after simplification.
This rule is somewhat conservative, as we will discuss in \autoref{sec:limitations-of-simplification}. 

The following excerpt of \fn{simplify} sketches a way in which we can write a simplification routine which respects these two rules.
\begin{align*}
  &\fn{simplify}~\term{keepIn}~(\Wrap{\term{id}}{\term{t}, \term{padding}}, \term{region})\\
  &\qquad=\fn{simplify}~(\fn{offsetPolygon}~(-\term{padding})~\term{keepIn})~\term{t} \\
  &\fn{simplify}~\term{keepIn}~(\Constr{JoinV}{t_1, t_2}, \term{region}) \\
  &\qquad=(\Constr{JoinV}{\fn{simplify}~\term{ps}_1~t_1, \fn{simplify}~\term{ps}_2~t_2}, \term{region})\\
  &\qquad\qquad\text{where}~\term{ps}_1 = \fn{simplifyPolygon}~p_1~p_2~\term{keepIn},\\
  &\qquad\qquad\phantom{\text{where}}~\term{ps}_2 = \fn{simplifyPolygon}~p_2~p_1~\term{keepIn},\\
  &\qquad\qquad\phantom{\text{where}}~p_1\phantom{s} = \fn{polygonOf}~(\fn{regionOf}~t_1), \text{and} \\
  &\qquad\qquad\phantom{\text{where}}~p_2\phantom{s} = \fn{polygonOf}~(\fn{regionOf}~t_2) \end{align*}
The \fn{simplify} operation takes two parameters: a \term{keepIn} polygon, serving as the boundary that child terms' boundaries ought to stay inside of, and the term itself.
Two of the interesting cases are shown.
When we encounter a \Unit{Wrap} node, the \term{keepIn} boundary is offset to account for the \Unit{Wrap} node's padding, for child terms should not encroach on the space allocated for this node's padding.
The case for \Unit{JoinV} is somewhat more involved.
We find unsimplified polygonal outlines for $t_1$ and $t_2$ ($p_1$, and $p_2$, respectively), and proceed to simplify them using \fn{simplifyPolygon}.
This function is the workhorse of the simplification algorithm; it accepts a polygon to be simplified, a polygon to stay out of, and a polygon to stay inside of.
When simplifying the outline of $t_1$, for example, the first call to \fn{simplifyPolygon} encodes that we ought not simplify so much that we encroach on our sibling, $t_2$, or our parent's outline, \term{keepIn}.
The case for \Unit{JoinH} is identical to \Unit{JoinV}, and the case for \Unit{Atom}s is straightforward, as there's no simplification to be done.
\begin{align*}
  &\fn{simplifyPolygon}~\term{p}~\term{keepOut}~\term{keepIn}\\
  &\qquad=\makebox[1.97in][l]{$p$} \text{when}~p = p'\\
  &\qquad=\makebox[1.97in][l]{$\fn{simplifyPolygon}~p'~\term{keepOut}~\term{keepIn}$} \text{otherwise}\\
  &\hspace{0.3in}\qquad\text{where}~p'=\fn{simplifyCorner}~\term{keepOut}~\term{keepIn}~p~\text{or,}\\
  &\hspace{0.3in}\qquad\phantom{\text{where}~p'=}~\fn{simplifyAntiKnob}~\term{keepOut}~\term{keepIn}~p~\text{or,}\\
  &\hspace{0.3in}\qquad\phantom{\text{where}~p'=}~p\text{, if no others apply}
\end{align*}

\begin{figure}[b]
  \includegraphics[width=3.3in]{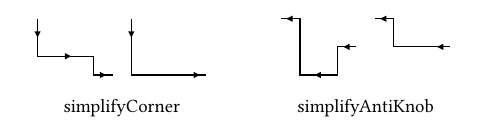}
  \caption{Example polygonal features, and simplifications.}
  \label{fig:simplify-polygon}
\end{figure}
 
\fn{simplifyPolygon} calls \fn{simplifyCorner} and \fn{simplifyAntiKnob} until a fixpoint.
\fn{simplifyCorner} and \fn{simplifyAntiKnob} are partial functions which perform the simplifications shown in \autoref{fig:simplify-polygon}.
They are partial not only because the simplification might not apply anywhere along the polygon, but because they also check that the simplification operation would not cause the outline to exceed \term{keepIn}, or intersect \term{keepOut}.

\subsection{Limitations of Simplification}
\label{sec:limitations-of-simplification}

\begin{figure}
  \includegraphics[width=3.3in]{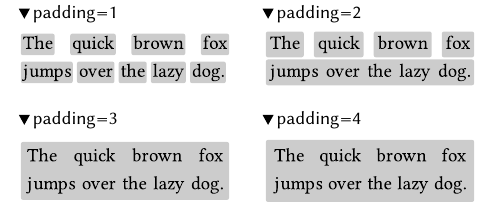}
  \caption{Simplifications with different padding values.}
  \label{fig:limitations}
\end{figure}
 
The definition of simplification relies on two rules; the first of which concerns constraints of parent-child relationships in the layout tree, and the second of which concerns constraints of sibling relationships.
In particular, the second rule is important for ensuring that if the regions of two siblings are disjoint prior to simplification, they will remain disjoint afterwards.
This rule is too conservative, and \autoref{fig:limitations} demonstrates why.
The layouts in \autoref{fig:limitations} all have the same layout tree.
Every fragment in the layout has the same \Unit{Wrap} node ancestor, so the entire layout could sensibly be represented by a single region.
And, when there is sufficient padding, this is exactly the result; the entire layout can be wrapped in a single outline representing the \Unit{Wrap}.
However, when insufficient padding is applied to the \Unit{Wrap} node (as in the top row of \autoref{fig:limitations}), simplification is unable to merge the regions into one.

Solving this problem in general is challenging.
It is difficult to know when two close, but disjoint, region outlines are compatible, \ie{}~they correspond to the same node in the layout tree, and are therefore permitted to share space.\footnote{
The reader may note that this is precisely the information that a \emph{region} surfaces, but simplification operates on region boundaries (polygons), not regions, and so we would need to attach some additional information to these boundaries.
}
Another challenge is finding a way to construct a bridge between the disjoint regions, in a way which avoids incompatible regions.

Intuitively, it seems like finding a \emph{convex hull} might be what we need.
The convex hull of a set of disjoint polygons is, after all, a well-behaved polygon which tightly wraps the disjoint set.
However, the shape of a rock is decidedly concave, and so some extra post-processing would be necessary to treat the resulting convex polygon so that it doesn't collide with its neighbors.
Additionally, there exists a notion of a \emph{rectilinear convex hull}~\cite{RectilinearConvexHull}.
Unfortunately, although this notion sounds promising, it does not closely match our visual expectation for well-formed regions.
Namely, in our formulation, the visual representation of a region cannot degenerate into a line segment (which does not encode information about the \emph{area} of some subtree).
Rectilinear convex hulls, in contrast, admit line segments.
Future work could address these challenges in order to construct outlines which are less prone to generating disjoint ``islands'' when the padding around a fragment is insufficient to merge with its compatible neighbors.

\section{Benchmarks}
\label{sec:benchmarks}

The layouts shown in this paper are generated from \rocks{}, our implementation of algorithms \Algop{L1}, \Algos{L1}, \Algos{L2a}, and \Algos{L2b}.
\rocks{} is implemented in about 5k LOC of TypeScript.
In addition to the rocks algorithms, our implementation contains \Boxes{}, a boxes-like layout algorithm, and \SBlock{}, a reimplementation of the layout algorithm in Hass~\cite{Hass}.

We ran our implementation on six representative source files, measuring differences between structured layouts---generated by \Algos{L1}, \SBlock{}, and \Boxes{}---as compared to unstyled, plain-text layouts.
The six source files (7.6k LOC in total) were collected from popular repositories on GitHub, with the exception of one example (layout), which was written by the first author.
The examples are written in three languages: TypeScript (core, diff-objs), Python (functional, simplex), and Haskell (solve, layout).
The examples also cover a range of coding styles, from imperative, statement-centric programming (core, functional, simplex), to highly nested, expression-centric programming (diff-objs, solve, layout).

\begin{figure}[b]
  \includegraphics[width=3.3in]{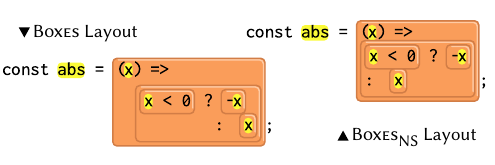}
  \caption{\Boxes{} vs \Boxesns{} for \Abs{}.}
  \label{fig:boxes-vs-boxes-ns}
\end{figure}
 
\newcommand{\rotheader}[2]{{#2}}

\newcommand{\pair}[2]{#1 (#2)
}

\newcommand{\xxx}{\rc{xxx}}

\newcommand{\pz}{\phantom{0}}

\newcommand{\columnAsCaption}[1]{\textbf{#1}}
\newcommand{\captionSpace}{\\[2pt]}

\setlength{\tabcolsep}{8pt}

\begin{figure*}
    \begin{subfigure}[t]{\textwidth}
  \centering
\begin{tabular}{ l || c c || c c }
    \letter{a}
    &
    &
    &
\multicolumn{2}{c}{\columnAsCaption{Running Time (\textit{s} (\%))}}
    \captionSpace
    &
    \rotheader{p{0.2in}}{LOC}
    &
{Fragments}
    &
    \rotheader{p{0.5in}}{\Algop{L1}}    &
    \rotheader{p{0.5in}}{\Algos{L1}}
    \\ \hline
      core.ts & 3020 &  20k & \phantom{<}8.2\pz{}\pz{} & \phantom{<}2.1\pz{}\pz{} (3.9$\times{}$)\pz{} \\
 diff-objs.ts &   25 & 0.2k &               <0.01\pz{} &               <0.01\pz{}     (0.19$\times$) \\
functional.py & 2233 & 5.3k &     \phantom{<}0.91\pz{} &     \phantom{<}0.25\pz{} (3.7$\times{}$)\pz{} \\
   simplex.py &  339 & 1.0k &         \phantom{<}0.032 &         \phantom{<}0.028 (1.2$\times{}$)\pz{} \\
     solve.hs & 1736 & 6.8k & \phantom{<}1.0\pz{}\pz{} &     \phantom{<}0.33\pz{} (3.1$\times{}$)\pz{} \\
    layout.hs &  285 & 2.0k &         \phantom{<}0.085 &         \phantom{<}0.064 (1.3$\times{}$)\pz{} \\
  \end{tabular}
\end{subfigure}
     
    \vspace{0.20in}
    \begin{subfigure}[t]{\textwidth}
  \centering
  \begin{tabular}{ l || c c c c c }
    \letter{b}
    &
\multicolumn{5}{c}{\columnAsCaption{Mean Line Width (\textit{px} (\%))}}
    \captionSpace
    &
    \rotheader{p{0.5in}}{Unstyled}      &
    \rotheader{p{0.5in}}{\Algos{L1}}    &
    \rotheader{p{0.5in}}{\SBlock{}}     &
    \rotheader{p{0.5in}}{\Boxesns{}}    &
    \rotheader{p{0.5in}}{\Boxes}
    \\ \hline
      core.ts & 260\phantom{.}     (0.78) & 310\phantom{.}     (0.91) & 310\phantom{.}     (0.91) & 340\phantom{.} (1.0) & 520\phantom{.} (1.5) \\
 diff-objs.ts & 280\phantom{.}     (0.38) & 340\phantom{.}     (0.46) & 340\phantom{.}     (0.46) & 740\phantom{.} (1.0) & 800\phantom{.} (1.1) \\
functional.py & 180\phantom{.}     (0.97) & 210\phantom{.} (1.1\pz{}) & 210\phantom{.} (1.1\pz{}) & 180\phantom{.} (1.0) & 350\phantom{.} (1.9) \\
   simplex.py & 180\phantom{.} (1.3\pz{}) & 210\phantom{.} (1.5\pz{}) & 210\phantom{.} (1.5\pz{}) & 140\phantom{.} (1.0) & 380\phantom{.} (2.7) \\
     solve.hs & 330\phantom{.}     (0.78) & 370\phantom{.}     (0.88) & 370\phantom{.}     (0.88) & 430\phantom{.} (1.0) & 610\phantom{.} (1.4) \\
    layout.hs & 270\phantom{.}     (0.73) & 300\phantom{.}     (0.84) & 300\phantom{.}     (0.84) & 360\phantom{.} (1.0) & 390\phantom{.} (1.1) \\
  \end{tabular}
\end{subfigure}
     
    \vspace{0.20in}
    \begin{subfigure}[t]{\textwidth}
  \centering
  \begin{tabular}{ l || c c c | c c c }
    \letter{c}
    &
\multicolumn{3}{c|}{\columnAsCaption{Mean Horizontal Mesh}} &
    \multicolumn{3}{c }{\columnAsCaption{Mean Vertical Mesh}}
    \\
    &
    \multicolumn{3}{c|}{\columnAsCaption{Distance w.r.t. Unstyled (\textit{px} (\%))}} &
    \multicolumn{3}{c }{\columnAsCaption{Distance w.r.t. Unstyled (\textit{px} (\%))}}
\captionSpace
    &
    \rotheader{p{0.5in}}{\Algos{L1}}    &
    \rotheader{p{0.5in}}{\SBlock{}}     &
    \rotheader{p{0.5in}|}{\Boxesns{}}   &
\rotheader{p{0.5in}}{\Algos{L1}}    &
    \rotheader{p{0.5in}}{\SBlock{}}     &
    \rotheader{p{0.5in}}{\Boxesns{}}
    \\ \hline
      core.ts & 4.8 (0.36) & 4.8 (0.36) & 13\phantom{.}\pz{} (1.0) &           \pz{}9.2 (0.52) & 21\phantom{.} (1.2) & 18\phantom{.}\pz{} (1.0) \\
 diff-objs.ts & 5.4 (0.13) & 5.4 (0.13) & 41\phantom{.}\pz{} (1.0) & 22\phantom{.}\pz{} (0.85) & 32\phantom{.} (1.2) & 26\phantom{.}\pz{} (1.0) \\
functional.py & 7.4 (0.33) & 7.4 (0.33) & 22\phantom{.}\pz{} (1.0) &           \pz{}6.2 (0.64) & 11\phantom{.} (1.1) &           \pz{}9.7 (1.0) \\
   simplex.py & 7.1 (0.88) & 7.1 (0.88) &           \pz{}8.0 (1.0) &           \pz{}7.7 (0.79) & 11\phantom{.} (1.1) &           \pz{}9.7 (1.0) \\
     solve.hs & 4.9 (0.24) & 4.9 (0.24) & 20\phantom{.}\pz{} (1.0) & 18\phantom{.}\pz{} (0.87) & 22\phantom{.} (1.1) & 20\phantom{.}\pz{} (1.0) \\
    layout.hs & 3.8 (0.33) & 3.8 (0.33) & 12\phantom{.}\pz{} (1.0) & 19\phantom{.}\pz{} (0.94) & 23\phantom{.} (1.1) & 20\phantom{.}\pz{} (1.0) \\
  \end{tabular}
\end{subfigure}
 \caption{
  Benchmarks.
Performance speedup of \Algos{L1} (average of 10 runs) is due to the use of stateful regions and corresponding optimization.
Line widths and mesh distances are reported in absolute terms as well as relative to \Boxesns{}.
}
\label{fig:evaluation}
\end{figure*}

Regarding the box-layout baseline for comparison, we actually implemented two strategies, \Boxes{} and \Boxesns{}.
The difference between these algorithms is only that \Boxes{} renders spacers, whereas \Boxesns{} ignores them.
\autoref{fig:boxes-vs-boxes-ns} shows their results on the \Abs{} example program.
In \Boxes{}, since \Unit{Newline}s are local to their subtree, whitespace cannot escape the nearest \Unit{Wrap} node.
Most implementations of boxes layout, therefore, ignore preceding whitespace, and let the padding applied to each \Unit{Wrap} node convey the nesting.\footnote{Compared to the \Boxes{} layout in \autoref{fig:boxes-vs-boxes-ns}, the \Boxes{} layout in \autoref{fig:blocks-vs-rocks} includes hand-tuned preceding whitespace to align the \texttt{:} and \texttt{?} characters.
It would be difficult to mechanically ensure alignment without a mechanism like that described in \autoref{sec:extensions-2a}.}

\autoref{fig:evaluation} summarizes our experiments: measuring the performance of our algorithm with pure and stateful regions, and measuring error with respect to unstyled layout in terms of both line- and fragment-based metrics.
We describe each of these three considerations below.

\subsection{Performance}

\autoref{fig:evaluation}~\letter{a} details the time taken to layout each example on \Algop{L1} and \Algos{L1}. In addition to stateful regions, our implementation of \Algos{L1} includes a \emph{spatial data structure} which accelerates calls to $\fn{leading}_\text{R}$.
We do not measure \Algos{L1} without this optimization because it would have similar performance to \Algop{L1}.
The key benefit of stateful regions is that it enables this optimization which leads to a significant speedup in most cases.
For small examples, the overhead of constructing a timetable prior to layout outweighs any gains from the spatial data structure, and \Algos{L1} is \emph{slower} than \Algop{L1}.
Apart from diff-objs, however, the optimizations in \Algos{L1} lead to it being between 1.2 and 3.9 times faster than \Algop{L1}.

\subsection{Line Width}

The ``Mean Line Width'' columns in \autoref{fig:evaluation}~\letter{b} report mean widths in pixels, as well as relative performance compared to \Boxesns{}, for the rendered lines pertaining to that benchmark. 

Among the structured layouts,
\rocks{} generates the narrowest lines for several benchmarks (core, diff-objs, solve, and layout) but not all;
\Boxesns{} produces narrower lines for the remaining benchmarks (functional and simplex), because it ignores indentation.
Ignoring indentation is especially helpful for examples in which indentation is the primary source of structure (\ie{}~in block-based languages such as TypeScript and Python).
In the expression-heavy TypeScript example (diff-objs) and the Haskell examples, algorithm \Algos{L1} produces shorter lines on average.

\Boxes{} is worse than \Boxesns{} (and \Algos{L1}) in every case, which is not surprising given that it does not have the advantage of ignoring preceding whitespace, and also cannot escape newlines as demonstrated in \autoref{fig:boxes-vs-boxes-ns}.
Given this sanity check, we do not report measurements for \Boxes{} in subsequent comparisons.

Notably, \SBlock{} produces layouts with identical mean line width to \rocks{} because its procedure for horizontally positioning fragments on a line is essentially the same as in \rocks{}:
they both place fragments as far to the left as possible, separating them by \fn{spaceBetween}, or one or more \emph{h-gadgets} as in Hass.
In both algorithms, the horizontal position of a fragment is solved \emph{before} its vertical position, so the latter cannot affect the former.
However, the \rocks{} and \SBlock{} algorithms differ in how they allocate vertical space, with
\SBlock{} layouts exhibiting higher error in the vertical placement of fragments (as described in more detail below).

The preceding compares the structured layouts to one another.
For evidence that the structured text layouts are not unreasonably larger than their unstyled, flat counterparts, \autoref{fig:evaluation} also reports mean line width for Unstyled layouts.
For each benchmark, the mean line width for \Algos{L1} is less than 21\% larger than this baseline.

\begin{figure}[b]
  \includegraphics[width=3.3in]{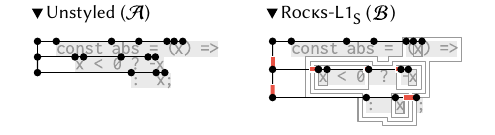}
  \caption{Mesh distance visualized for Unstyled and \Algos{L1} layouts. Red bars visualize the difference in length between corresponding segments in layouts $\mathcal{A}$ and $\mathcal{B}$. The mesh distance between $\mathcal{A}$ and $\mathcal{B}$ is the sum of the lengths of red bars.}
  \label{fig:mesh-distance}
\end{figure}
 
\begin{figure*}[t]
  \includegraphics[width=7in]{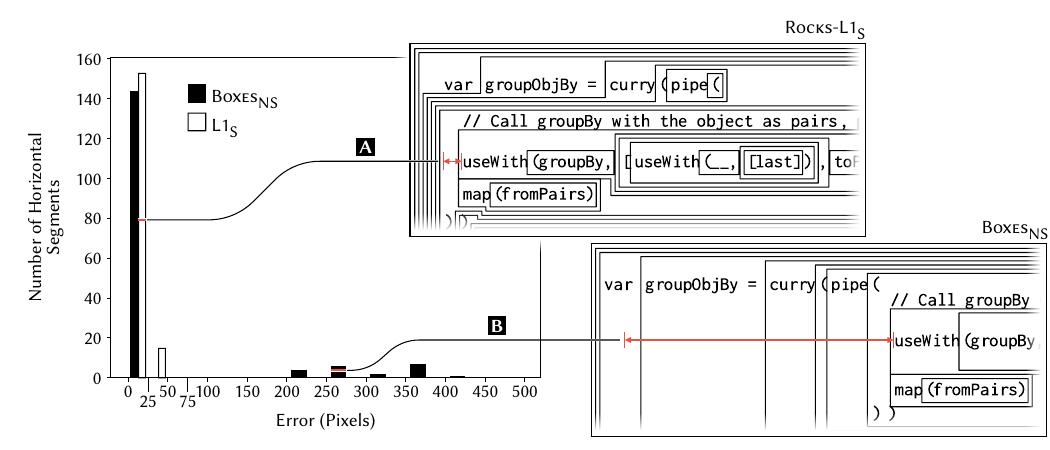}
  \caption{Distribution of error in horizontal segments for the diff-objs benchmark.}
  \label{fig:histogram}
\end{figure*}
 
\subsection{Measuring Layout Error}

Line width is important, but judging two layouts simply by comparing line widths would not tell the whole story---only the first and last fragments on a line contribute to its width.
How can we compare layouts in a fine-grained way that accounts for \emph{all} fragments?

We devise a new metric, called \emph{mesh distance}, which we believe to be a fair way to judge the amount of \emph{stretching} or \emph{error} between two text layouts $\mathcal{A}$ and $\mathcal{B}$.
This notion has two components; one which measures the vertical segments, and one which measures the horizontal.
We require layouts $\mathcal{A}$ and $\mathcal{B}$ to have the same number of vertical and horizontal segments.
Indeed, the mesh distance is a measure between two layouts which differ only in the \emph{position} of their constituent rectangles.
The mesh distance is undefined for layouts which represent different text, or the same text but with differently chosen line breaks.

\autoref{fig:mesh-distance} visually summarizes the mesh distance between an Unstyled and \Algos{L1} layout.
The mesh distance is calculated by drawing horizontal segments connecting each adjacent fragment on a line, and vertical segments connecting the first fragment on adjacent lines.
The length of these segments are compared pairwise between layouts to derive a notion of error between a ground-truth layout ($\mathcal{A}$) and a test layout ($\mathcal{B}$).
A more formal description of mesh distance is given in \refAppendixMeshDistance{}.

\paragraph{Comparing Mesh Distances}

For each of the benchmarks in \autoref{fig:evaluation}, we calculated the mesh distance between a layout with no padding applied anywhere (implemented by \Algos{L1}) and a layout with uniform padding applied to every node in the input tree.
If no padding is applied anywhere, then \Algos{L1} degrades to ordinary flat text layout as one might expect from an ordinary text editor.
Thus, it serves as our ground truth.
Of course, an effective structured text layout must necessarily introduce \emph{some} error from this unstructured ground truth; the unstructured layout has no space between boxes by which to show structure.
Accordingly, we are not interested in layouts with zero error, only in minimizing the layout's error, as measured by the mesh distance.
The mesh distances in \autoref{fig:evaluation}~\letter{c} are normalized to \Boxesns{}, since \Boxesns{} is the most commonly used structured text visualization in our experience.

\autoref{fig:evaluation}~\letter{c} shows that algorithm \Algos{L1} does indeed produce layouts that are more similar to the reference plain-text layouts than our \SBlock{} and \Boxesns{} reference implementations.
In the area of horizontal error, \Algos{L1} and \SBlock{} produce identical error because they agree on the horizontal position of every fragment.
For all benchmarks, these algorithms produce less horizontal error than \Boxesns{}, but the degree of improvement over \Boxesns{} varies.
In particular, we notice that for examples and languages which are more expression-oriented (diff-objs, solve, layout), \Algos{L1} produces \emph{significantly} more compact layouts, whereas the difference is not as pronounced for examples which demonstrate a more statement-oriented programming style (core, functional, simplex).

Considering the vertical mesh distance, we note that \Algos{L1} produces layouts with the \emph{least} vertical error in every benchmark, with \SBlock{} producing layouts with the \emph{most} vertical error.
This is unsurprising given the key advantage is that a rock from a previous line can cross the line boundary into subsequent lines if there is no vertical overlap, while fragments (and their associated borders) in an s-block are constrained to reside entirely within a single line.

\paragraph{Case Study: diff-objs}

In those examples which use an expression-oriented style, it is often the case that in deeply nested expressions, the \Boxesns{} layout places its output rectangles very far from the reference.
One such example is demonstrated in \autoref{fig:histogram}, which shows a fragment of the rendered output of the diff-objs benchmark using the \Algos{L1} and \Boxesns{} algorithms.

For this program, both layouts contain many horizontal segments with small error, but \Boxesns{} contains a few horizontal segments with very large error.
Compare, for example, the error due to the fragment \texttt{useWith}, which is small in \Algos{L1} (see \autoref{fig:histogram}~\letter{a}) but large in \Boxesns{} (see \autoref{fig:histogram}~\letter{b}).
Since the {entire} subexpression of a call is required to fit inside a box in \Boxesns{}, function arguments which span multiple lines are often pushed much further rightward than in the unformatted text.
Algorithm \Algos{L1} avoids this problem by permitting the boundary of a subexpression to flow into the shape of an arbitrary polygon, and allowing newlines to ``cut'' through the entire layout tree.

Of course, algorithm \Algos{L1}'s more flexible outlines come with tradeoffs.
The layout produced by \Algos{L1} is arguably harder to read since the more complex outlines are harder to visually parse.
Furthermore, it is possible for a subexpression to be represented as two or more disjoint polygons in \Algos{L1} (see the application of the function \texttt{pipe} in \autoref{fig:histogram}, for example), a situation that cannot occur in \Boxesns{} (see also \nameref{sec:limitations-of-simplification}).

\subsection{Smoothness}

The experiments reported in \autoref{fig:evaluation} do not invoke simplification, and we don't report the performance impact of this post-processing step.
In addition to performance impact, it could be interesting to count the number of corners reduced by simplification, but this is a task left for future work.

Regardless, judging the quality and preferences of the polygonal outlines resists quantitative treatment as we've done in evaluating compactness.
The various examples given throughout the paper,
as well as additional ones in \refAppendixApplications{},
indicate the nature and variety of polygonal outlines generated by algorithms \Algos{L1}, \Algos{L2a}, and \Algos{L2b}, with and without simplification.
But ultimately, these will need to be evaluated by users; an important direction for future work.
 
\section{Conclusion and Future Work}
\label{sec:disco}

In this paper, we identified arbitrary rectilinear polygons---which we call ragged blocks, or rocks---as a building block for rendering nested text visualizations, and detailed a family of algorithms for computing compact rock layouts.
On a set of benchmark source code files, we showed that the layouts produced by our implementation, \rocks{}, are significantly more compact than traditional layout techniques involving only rectangles, or boxes, further improving upon recent techniques in the literature involving s-blocks.
This result is a step towards more seamlessly integrating text-based and structure-based editing into IDEs for programming.

There are many natural ways to build on the techniques in this paper.
One is to continue optimizing the implementation of the layout algorithms, to make them more practical for large files.
Another is to investigate avenues for additional simplification, as discussed.

Furthermore, while the algorithms themselves are generally language agnostic (which is how we could easily evaluate on benchmarks written in different languages), a more full-featured editor for a specific language will require fine-tuned pre-processing to convert syntax trees in that language to appropriately configured layout trees.
For example, the general front-end parser we currently use in \rocks{} sometimes generates incorrect Haskell parse trees; these would need to be refined before subsequent fine-grained styling and rendering.

On a related note, while algorithm \Algos{L2a} supports column constraints, it may be impractical to expect users to manually write them.
Instead, general language-specific conventions could be used to inform default column constraints---for example, definitions within a particular block could be implicitly constrained for alignment, whereas definitions across top-level blocks would not.

Beyond these algorithmic concerns is the need to study how ragged blocks might facilitate useful interactions as compared to existing mechanisms (namely, boxes and s-blocks).
In \refAppendixApplications{}, we present a small gallery of examples that, compared to visualizing all blocks, include finer-grained stylistic choices for specific tasks:
highlighting differences between two versions of a document, as well as
conveying error messages and other semantic information about its content.
These examples involve both programming- and natural-language documents, hinting at how future code editors and document-rendering GUIs could employ domain-specific styling of ragged blocks as a rich channel for communicating information to users.
Future work should more thoroughly explore these and other applications.

Another direction for future work is to reexamine existing text visualizations in light of the new layout capabilities, and to imagine new visualizations that they enable.
Compared to the well-studied realm of program visualizations, the space of non-code text visualizations is rather less explored.
Perhaps this is because prose is not as highly structured as code, and because users do not expect the same degree of richness from ``common'' tools such as word processors and web browsers as from coding tools.
But we also believe the lack of expressive layout techniques has been a limiting factor for experimentation in GUIs for text-heavy productivity tools.

Ultimately, the motivation for this work is to enable new text visualizations that users find helpful as they read and write a variety of documents.
So, in addition to the quantitative benchmarking we have pursued to evaluate our techniques so far, it will be crucial to understand how different design tradeoffs---between compactness, smoothness, and the use of various decorations---are perceived by users in different usage scenarios.


\clearpage
\appendix

\renewcommand{\leftmark}{}
\renewcommand{\rightmark}{}
\pagestyle{headings}

\section{Layout with Column Constraints}
\label{sec:version-2a}

In this section, we will outline modest changes to algorithm \Algop{L1} which will permit columns to remain aligned after layout.
The key insight is this: in \Algop{L1}, given two regions, call them $a$ and $b$, \joinh{} placed $b$ directly after $a$, translating it by $a$'s advance.
Instead of translating immediately, we could impose a \emph{constraint} on $b$, permitting its horizontal position to be \emph{no less than} the position of $a$ plus $a$'s advance.
We might assign $b$'s horizontal position to a variable, say $p_0$.
Then, using $p_0$ to describe the horizontal position of fragments in other lines has the effect of constraining those fragments to be horizontally aligned with $b$.

In order to model the addition of constrained rectangles, we introduce a new kind of term into our grammar of layout trees:

\vspace{0.1in} 

\begin{tabular}{l c l}
    $t$ & \pdn{} &
    $\Constr{Atom}{x_i}$
    | \new{$\Pin{\term{pid}}{x_i}$}
    | $\Constr{Spacer}{w_i}$ | \\ &&
      $\Wrap{\term{id}}{t, \term{padding}}$
| $\Constr{JoinV}{t, t}$
    | $\Constr{JoinH}{t, t}$ \\[4pt]

$\term{layout}$ & \pdn{} & [\new{$\List{(\term{stack}, \term{pid}, \mathbf{v})}$}]
  \end{tabular}

\vspace{0.1in} 

A \Unit{Pin} is a fragment, $x_i$, with an identifier, \term{pid}, which is a name uniquely identifying the constraint variable associated with the \Unit{Pin}.
We encode that two rectangles are constrained to the same horizontal position by wrapping them in the \Unit{Pin} constructor, and giving them the same $\term{pid}$, thus ensuring that their horizontal positions are determined by the same variable.

The type of layouts has also changed.
In algorithm \Algop{L1}, lines were the atomic units with which layout was concerned---it was never necessary to peek inside a line---only to take the \fn{union} of partial lines, or \fn{translate} lines.
Not so for algorithm \Algop{L2a}.
The horizontal position of each individual \Unit{Stack} in the layout is determined by the result of constraint resolution, and so it is necessary to be able to iterate through these \Unit{Stack}s one-by-one.\footnote{In practice, it is usually the case that only a few \Unit{Stack}s in the layout are constrained, while the others are laid out exactly like in algorithm \Algop{L1}.
It is possible, then, to assign a variable to each \emph{contiguous group} of unconstrained stacks, rather than each stack.
This reduces the number of variables in the constraint problem, and has no effect on the resulting layout.
}
This is reflected in the type of an \Algop{L2a} layout, which is a nested list of \Unit{Stack}s, annotated with the $\term{pid}$ of the variable which determines the \Unit{Stack}'s horizontal position, and the nominal advance of the \Unit{Stack}, which serves the same role as the advance in algorithm \Algop{L1}.

\begin{figure}
  \includegraphics[width=3.3in]{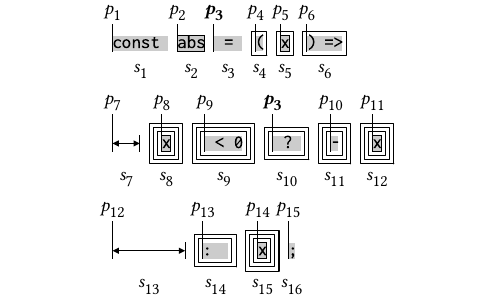}
  \begin{tabular}{r c l}
    $p_1$ & $\geq$ & $0$ \\
    $p_2$ & $\geq$ & $p_1 + \fn{advance}~s_1~s_2$ \\
    \multicolumn{3}{c}{$\dots$} \\
    $p_6$ & $\geq$ & $p_5 + \fn{advance}~s_5~s_6$ \\
    $p_7$ & $\geq$ & $0$ \\
    $p_8$ & $\geq$ & $p_7 + \fn{advance}~s_7~s_8$ \\
    \multicolumn{3}{c}{$\dots$} \\
    $p_{11}$ & $\geq$ & $p_{10} + \fn{advance}~s_{11}~s_{12}$ \\
    $p_{12}$ & $\geq$ & $0$ \\
    $p_{13}$ & $\geq$ & $p_{12} + \fn{advance}~s_{13}~s_{14}$ \\
    \multicolumn{3}{c}{$\dots$} \\
    $p_{15}$ & $\geq$ & $p_{14} + \fn{advance}~s_{15}~s_{16}$ \\
  \end{tabular}
  \begin{equation*}
    \text{Minimizing}~\sum_{i=1}^{15}p_i
  \end{equation*}

  \caption{Example constraint problem.}
  \label{fig:L2ap-constraint-problem}
\end{figure}

\autoref{fig:L2ap-constraint-problem} gives an example of the constraints we might construct for the \Abs{} program.
Each constraint takes the form of a linear inequality, where the first fragment on each line is constrained to be on the left margin.
A stack $s_i$ is \emph{at least} as far right as the stack to its left ($s_{i-1}$), plus the \fn{advance} between $s_{i-1}$ and $s_i$.
By giving an objective function which attempts to minimize the sum of the $p_i$s, we may phrase this problem as a linear program, and solve it automatically during layout.

\begin{figure}
  \includegraphics[width=1.5in]{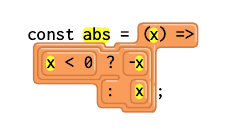}
  \caption{Example \Algop{2a} layout.}
  \label{fig:L2ap-result}
\end{figure}

Once the horizontal position of each \Unit{Stack} has been determined, each line is merged into a single region using $\fn{leading}_R$ just as in algorithm \Algop{L1}.
The result of using algorithm \Algop{L2a} instead of algorithm \Algop{L1} on the \Abs{} program yields the result shown in \autoref{fig:L2ap-result}.
A single constraint has been introduced, causing the \Unit{Stack}s corresponding to \texttt{?} and \texttt{:} to be aligned in the finished layout.

\clearpage
\section{Layout with Automatic Line Breaks}
\label{sec:version-2b}

\begin{figure}[b]
  \includegraphics[width=3.3in]{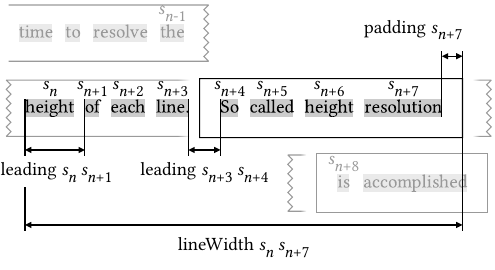}
  \caption{Example input tape.}
  \label{fig:L2bp}
\end{figure}
 
In algorithm \Algop{L2a}, we showed that adjacent stacks need not be placed tightly against one another.
Indeed, we can use $\fn{advance}_R$ as a kind of \emph{minimum bound} on the spacing between two stacks.
This minimum bound information is enough to implement a variety of text layout algorithms, and not only those which are suited for layout of pre-formatted text.
Algorithm \Algop{L2b}, as opposed to algorithms \Algop{L1} and \Algop{L2a}, takes \emph{unformatted} text as input.
That is, text which does not contain explicit newlines.
For this reason, we will refer to the linear stream of \Unit{Stack}s that \Algop{L2b} takes as input as the ``tape.''\footnote{Sometimes called ``idiot tape:'' a stream of text for layout with no additional instructions on \emph{how} it should be laid out.}
This is reflected in the new grammar of layout trees for \Algop{L2b}, which dispenses with the \Unit{JoinV} constructor for concatenating two text layouts vertically:

\vspace{0.0in} 

\begin{center}

\begin{tabular}{l c l}
    $t$ & \pdn{} &
    $\Constr{Atom}{x_i}$
    | $\Constr{Spacer}{w_i}$ | \\ &&
      $\Wrap{\term{id}}{t, \term{padding}}$
    \removed{| $\Constr{JoinV}{t, t}$}
    | $\Constr{JoinH}{t, t}$ \\[4pt]

$\term{layout}$ & \pdn{} & [\new{$\List{(\term{stack}, \mathbf{v})}$}]
  \end{tabular}

\end{center}

\vspace{0.1in} 

Thus, algorithm \Algop{L2b}'s job is twofold: it must find good locations for line breaks, and then having inserted those breaks, it must find the positions of each rectangle, $x_i$, in the fashion of algorithms \Algop{L1} and \Algop{L2a}.
In order to accomplish the former goal, we have re-implemented a subset of the algorithm given by \citeauthor{BreakingParagraphs}, modifying it to use \Unit{Stack}s in place of rectangles, thus allowing our implementation to support structured text.

\citeauthor{BreakingParagraphs}' algorithm works by iterating once over the input tape, adding and removing items from a set, called the \emph{feasible set}, describing the set of possible line breaks in the layout.
On each element in the tape (the current point), a new line break can be added to the set if it is deemed feasible (there exists another line break in the feasible set which, when paired with the current point, would make a feasible line), and some breaks can be removed from the set if they are no longer feasible (there \emph{does not} exist a line break in the feasible set which could form a valid line with the current point).
Each line break in the feasible set remembers the history of line breaks before them, and the cost of this history trace, according to an objective function.
Once the entire tape has been traversed, the trace with the lowest cost is selected, and the text is broken according to the breaks in the optimal trace.

The advance (equivalent to \emph{width} in the unstructured algorithm) of an atomic fragment of text is considered only when determining if a subset of the input tape consistitutes a feasible line.
This feasibility test compares the ``ideal'' line width (an input to the algorithm) to the line's actual width, and calculates a cost based on the difference between these quantities.
Therefore, in order to modify the algorithm to support structured text, we need only update the function which calculates the length of a line to consider the presence of padding between fragments.

We define the following function, \fn{lineWidth}, which calculates the length of line beginning at stack $s_i$ and ending at stack $s_j$.

\begin{align*}
  &\fn{lineWidth}~s_i~s_j=\Sigma_{p_{s_i}} + \Sigma_{p_{s_ij}} + \Sigma_a\\
  &\qquad\text{where }\\
  &\qquad\qquad\Constr{Stack}{\Sigma_{p_{s_i}}, \term{cells}_{s_i}} = s_i, \text{and}\\
  &\qquad\qquad\Constr{Stack}{\Sigma_{p_{s_j}}, \term{cells}_{s_j}} = s_j, \text{and}\\
  &\qquad\qquad\Sigma_a = \sum_{\mathclap{(a, b) \in \mathcal{P}}}\fn{advance}~a~b\\
  &\qquad\qquad\text{where }\\
  &\qquad\qquad\qquad \mathcal{P} = \text{adjacent pairs, $(a, b)$, in $s_i, s_{i+1}, \dots, s_{j}$}
\end{align*}

Note that special care is taken to include \emph{all} of the padding surrounding the first and last \Unit{Stack}s on the candidate line, since these \Unit{Stack}s cannot possibly engage in space sharing with their left and right neighbors, respectively. (For example, in \autoref{fig:L2bp}, \Unit{Stack} $s_n$ cannot share space with \Unit{Stack} $s_{n-1}$, and \Unit{Stack} $s_{n+7}$ cannot share space with \Unit{Stack} $s_{n+8}$, despite the fact that they are both wrapped by the same \Unit{Node}\footnote{Note, however, that these \Unit{Stack}s \emph{could} (and will) participate in space sharing as their lines are merged into a single region.
}).

\begin{figure}
  \includegraphics[width=3.3in]{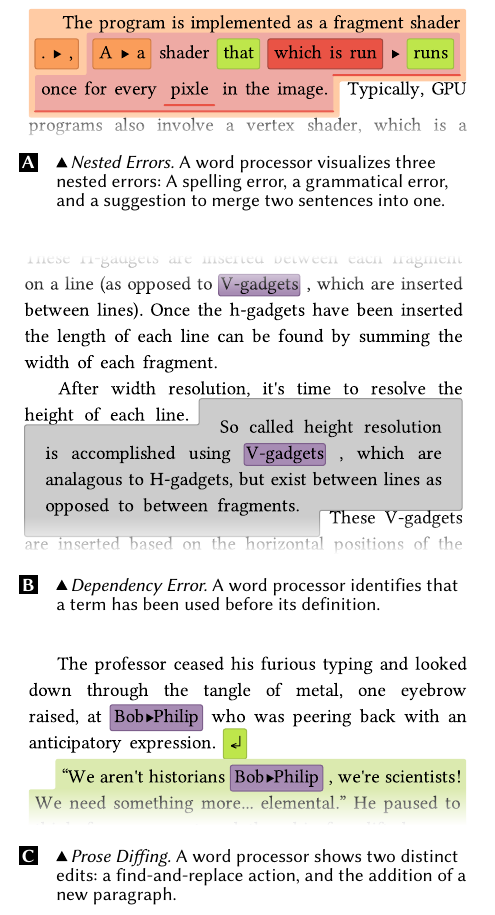}
  \caption{Example prose visualizations.}
  \label{fig:prose-gallery}
\end{figure}

Once the location of line breaks are known, the input layout tree may be modified by judiciously inserting vertical concatenation constructors (\Unit{JoinV}) according the results of the search.
Then, algorithm \Algop{L1}, for example, could be used to find the final position of each rectangle.
In our implementation, a separate but similar algorithm is used which \emph{fully justifies} the \Unit{Stack}s on each line, evenly spacing them so the leftmost \Unit{Stack} on the line abuts the left margin, and the rightmost \Unit{Stack} abuts the right margin. (See the examples in \autoref{fig:prose-gallery}, for example.)
 
\clearpage
\section{Examples}
\label{sec:applications}

\begin{figure*}
  \includegraphics[width=7in]{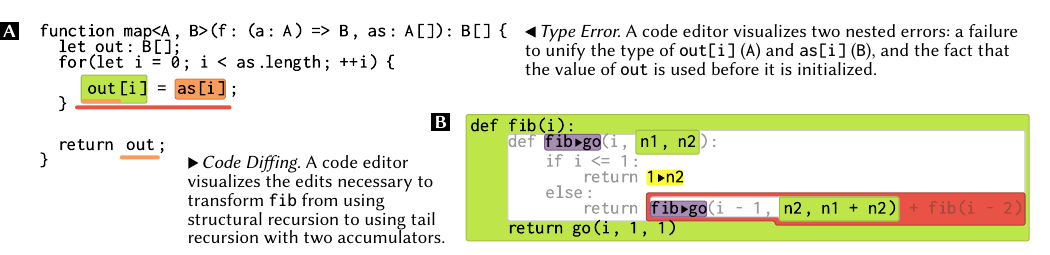}
  \caption{Example code visualizations.}
  \label{fig:code-gallery}
\end{figure*}
 
The predominant use case of rocks depicted throughout the paper, and evaluated in our benchmarks, is to systematically render all nested substructures uniformly with border and padding.

In this section, we present a small gallery of examples that include finer-grained stylistic choices for specific tasks:
highlighting differences between two versions of document, as well as
conveying error messages and other semantic information about its content.
These examples involve both programming- and natural-language documents, hinting at how future code editors and document-rendering GUIs could employ domain-specific styling of ragged blocks as a rich channel for communicating information to users.

\subsection{Code}
\label{sec:applications-code}

\autoref{fig:code-gallery} demonstrates how structured text could be applied to the task of visualizing code errors and diffs.

\paragraph{Visualizing Errors in Code}
The code fragment in \autoref{fig:code-gallery} \letter{a} shows an example of nested errors in TypeScript.
In the example, the programmer forgot to apply the function \texttt{f} to the elements of the input list \texttt{as}, causing \texttt{out} to be assigned to a value of the wrong type.
Additionally, the programmer failed to initialize \texttt{out}, and that error is \emph{also} surfaced at the point of assignment.

A structured text visualization can show overlapping error contexts, such as the context for the type error which is indicated with orange underlines, as opposed to the context for the initialization error, which is shown with red underlines.
The green and orange boxes mark the two expressions with incompatible types.

\paragraph{Code Diffs}
In example \autoref{fig:code-gallery} \letter{b}, a Python definition of a \texttt{fib} function written in ``natural'' recursive style is transformed into a tail-recursive version which uses two accumulator parameters.
The original definition is wrapped in a new function (the outermost green box), new parameters are added, and the inner definition of \texttt{fib} is renamed to \texttt{go} (purple box).
Structured diffs, like that shown, are an alternative to traditional line-based diffs.
Because a structural diff is aware of the tree underlying the source program, it can render a change as a sequence of more meaningful \emph{semantic} changes to the program, such as the wrapping of one function in another.

\newpage

\subsection{Prose}
\label{sec:applications-prose}

Analogous to the previous examples, \autoref{fig:prose-gallery} includes examples of visualizing errors and differences in natural-language documents.

\paragraph{Visualizing Errors in Prose}

In the face of an error depicted in \autoref{fig:prose-gallery} \letter{a},
the editor provides suggestions for how the user might modify the text.
But instead of presenting them individually, an editor endowed with a structured text visualization can present all options simultaneously, so that the user can view and apply them in any order they see fit.

Grammar and spelling checkers are ubiquitous in writing tools, but we might also imagine a different, \emph{semantic} analysis which ensures that terms are not used before their definition.
\autoref{fig:prose-gallery} \letter{b} shows a visualization for this hypothetical analysis, which shows a term, in purple, that is used before its definition, and its definition which is marked in gray.

\paragraph{Prose Diffs}
\autoref{fig:prose-gallery} \letter{c} shows two unrelated, but overlapping modifications to the text.
Nesting permits the visualization to imply the ordering of the edits, despite the fact that they overlap.

\subsection{Formulas}

\begin{figure}[b]
\includegraphics[width=3.3in]{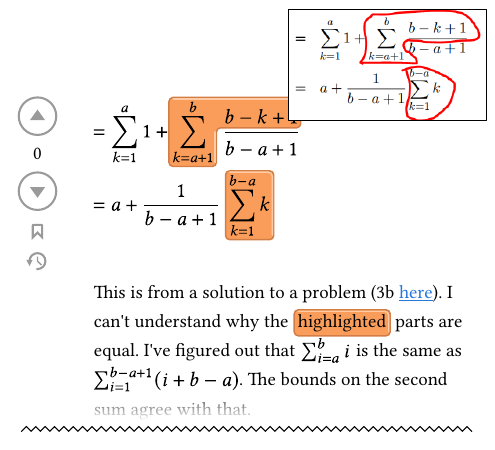}
\caption{Example formula visualization. Mockup of a Stack Exchange post using \rocks{} to annotate parts of an equation. The original image accompanying the post is shown in the upper right. (\url{https://math.stackexchange.com/questions/2624345/stuck-with-a-sum-problem-change-of-lower-bound})}
\label{fig:sum-problem}
\end{figure}

More structured than prose, but less so than code, are math formulas.
\autoref{fig:sum-problem} imagines an extension to Stack Exchange whereby \rocks{} could be used to annotate parts of an equation to facilitate discussion.
The example given in \autoref{fig:sum-problem} is based on a post where contributors were confused about the extent and meaning of the author's hand-drawn annotations (shown in the top-right of \autoref{fig:sum-problem}).
\rocks{} could offer a mechanism for contributors to be more precise when interacting in asynchronous, online contexts.
 
\clearpage
\onecolumn

\section{\Abs{} Timetable}
\label{sec:timetable-abs}

\begin{figure}[h]
  \includegraphics[width=7in]{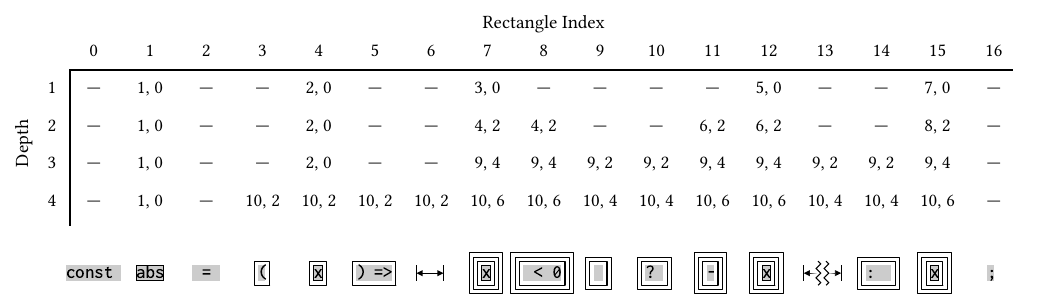}
\caption{Timetable corresponding to the \Abs{} example.
For brevity, \Unit{Cell} constructors are omitted; each pair represents the \Unit{Cell} \term{id} and cumulative padding, respectively.
A visual representation of each fragment (and all of its wraps) is shown beneath its corresponding column.
}
\end{figure}
 
\twocolumn
 
\section{Mesh Distance}
\label{sec:mesh-distance}

Formally, we define the mesh distance as follows.
Let $\mathcal{A}$ be a text layout.
In keeping with our definition of the layout problem in \autoref{sec:version-1}, we say that a text layout is represented by the position of its constituent fragments (we don't evaluate the polygonal outlines when computing the mesh distance, so they can be ignored).
So, let $\List{x^{\mathcal{A}}_1, \dots, x^{\mathcal{A}}_n} \in \mathcal{A}$ be the laid-out fragments of layout $\mathcal{A}$.

We can further subdivide the set of fragments in each layout by their line.
We might say that in $\mathcal{A}$, for example:
\[
\List{x^{\mathcal{A}}_1, \dots, x^{\mathcal{A}}_n} = \List{x^{\mathcal{A}}_1, \dots\, x^{\mathcal{A}}_{l_1}} + \cdots + \List{x^{\mathcal{A}}_{l_{m-1} + 1}, \dots\, x^{\mathcal{A}}_{l_m}}
\]
Here, the layout has $m$ lines, $x^{\mathcal{A}}_1, \dots\, x^{\mathcal{A}}_{l_1}$ occuring on the first line, $x^{\mathcal{A}}_{l_1+1}, \dots\, x^{\mathcal{A}}_{l_2}$ occuring on the second, and so on.

For each fragment, we choose a representative point.
We'll choose the upper-left corner of each fragment as our canonical point, and define a function, $\fn{ul}~x$ to retrive this point.
For each line in $\mathcal{A}$, we construct a set of segments which connect every adjacent pair of fragments.
So, the first line of $\mathcal{A}$ yields $l_1 - 1$ segments:
\[
\mathcal{A}_\text{H} = \List{(\fn{ul}~x_1^{\mathcal{A}}, \fn{ul}~x_2^{\mathcal{A}}), \dots, (\fn{ul}~x_{l_1 - 1}^{\mathcal{A}}, \fn{ul}~x_{l_1}^{\mathcal{A}})}.
\]
We call these the \emph{horizontal segments} of the layout $\mathcal{A}$.

In a similar fashion, we can construct the \emph{vertical segments} of layout $\mathcal{A}$.
The vertical segments of $\mathcal{A}$ connect the first fragment on each line:
\[
\mathcal{A}_{\text{V}} = \List{(\fn{ul}~x_1^{\mathcal{A}}, \fn{ul}~x_{l_1+1}^{\mathcal{A}}), \dots, (\fn{ul}~x_{l_{m-2}+1}^{\mathcal{A}}, \fn{ul}~x_{l_{m-1}+1}^{\mathcal{A}})}.
\]

Then, the mesh distance of two layouts, $\mathcal{A}$ and $\mathcal{B}$, is defined as follows:
\begin{align*}
  \fn{meshDistance}_\text{H}~\mathcal{A}~\mathcal{B} &=\sum_{\mathclap{i \in |\mathcal{A}_\text{H}| = |\mathcal{B}_\text{H}|}} \fn{length}~\mathcal{B}_{\text{H}_i} - \fn{length}~\mathcal{A}_{\text{H}_i}~\text{and,}\\
  \fn{meshDistance}_\text{V}~\mathcal{A}~\mathcal{B} &=\sum_{\mathclap{i \in |\mathcal{A}_\text{V}| = |\mathcal{B}_\text{V}|}} \fn{length}~\mathcal{B}_{\text{V}_i} - \fn{length}~\mathcal{A}_{\text{V}_i}.
\end{align*}
 
\newpage \section{Boulder}

\begin{figure}[h]
\includegraphics{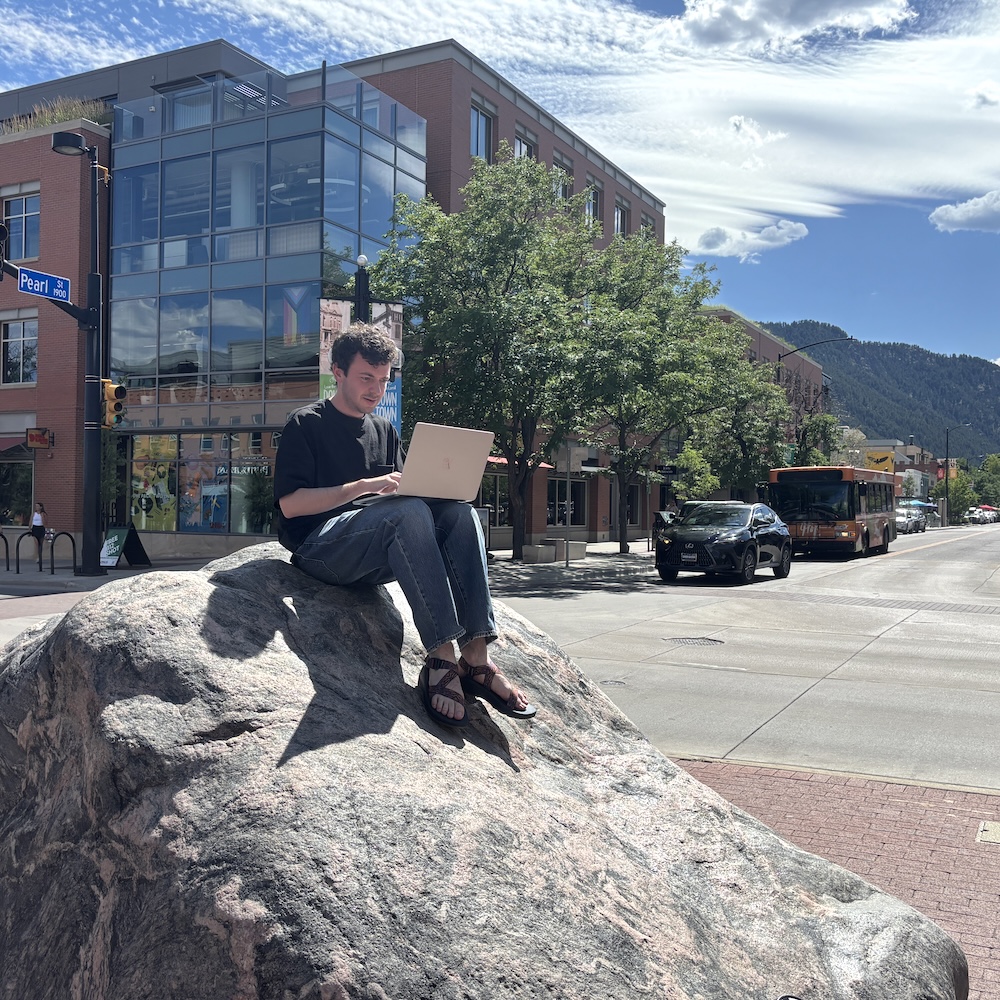}
\caption{Scaling the last boulder---July 9, 2025 (Boulder, CO).}
\end{figure}
 
\onecolumn

\section{Additional Benchmarking}

\paragraph{Stateful Regions}

Compared to the implementation of \Algos{L1} benchmarked in \autoref{sec:benchmarks}, the version of \Algos{L1} benchmarked in \autoref{fig:evaluation-performance-updated} has been further optimized.
The primary bottleneck of layout is $\fn{leading}_\text{R}$, which, as written in \autoref{fig:L1p}, requires $O(d n^2)$ comparisons where $d$ is the depth of the layout tree and $n$ is the number of fragments.
Using stateful regions, it is possible to significantly improve this by re-writing $\fn{leading}_\text{R}$ so that it tests \emph{lower} fragments in $\textit{region}_a$ first, and so can exit early when we've found a collision between fragments in $\textit{region}_a$ and $\textit{region}_b$.
Still, the algorithm spends much of its time in this inner loop, and care was taken to streamline the code within this path.

The primary operation of the inner loop in $\fn{leading}_\text{R}$ is $\fn{spaceBetween}$, which, in the stateful region implementation, traverses the timetable.
The timetable (described in \autoref{sec:stateful-regions}) is queried to test if two fragments might collide.
Because $\fn{spaceBetween}$ is invoked many times during layout, making these timetable queries fast can have an outsize effect on performance.
In particular, if the height of the table is large, and the fragments being tested are both deeply nested, we can waste time shedding common layers until we reach two table cells which derive from different $\textsf{Wrap}$ nodes.
To save time during traversal, we insert indirections or ``jumps'' into the table pointing to the next interesting $\textsf{Wrap}$, rather than repeating the topmost element.
(In \autoref{fig:timetable-abs}, for example, columns 1 and 4 would benefit from such an optimization).
When traversing the table, when an indirection is encountered, it is followed directly to the relevant cell, saving time.

These changes lead to the performance improvements shown in \autoref{fig:evaluation-performance-updated}.

\newcommand{\highlightNumber}[1]{\color{purple}#1}

\begin{figure*}[h]
\begin{subfigure}[t]{\textwidth}
  \centering
  \begin{tabular}{ l || c c || c c }
    &
    &
    &
    \multicolumn{2}{c}{\columnAsCaption{Running Time (\textit{s} (\%))}}
    \captionSpace
    &
    \rotheader{p{0.2in}}{LOC}
    &
    {Fragments}
    &
    \rotheader{p{0.5in}}{\Algop{L1}}    &
    \rotheader{p{0.5in}}{\Algos{L1}}
    \\ \hline
      core.ts & 3020 &          20k & \phantom{<}8.3\pz{}\pz{} & \highlightNumber{\phantom{<}0.22\pz{} (38 $\times{}$)} \\
 diff-objs.ts &   25 &         0.2k &               <0.01\pz{} & \highlightNumber{          <0.01\pz{} (\pz{}1 $\times{}$)} \\
functional.py & 2233 &         5.3k &     \phantom{<}0.93\pz{} & \highlightNumber{    \phantom{<}0.043 (21 $\times{}$)} \\
   simplex.py &  339 &         1.0k &         \phantom{<}0.033 & \highlightNumber{          <0.01\pz{} (\pz{}5 $\times{}$)} \\
     solve.hs & 1736 &         6.8k & \phantom{<}1.0\pz{}\pz{} & \highlightNumber{\phantom{<}0.05\pz{} (21 $\times{}$)} \\
    layout.hs &  285 &         2.0k &         \phantom{<}0.086 & \highlightNumber{    \phantom{<}0.011 (\pz{}7 $\times{}$)} \\
  \end{tabular}
\end{subfigure}
\caption{
  Updated running times for \Algos{L1} (cf. \autoref{fig:evaluation} \letter{a}).
}
\label{fig:evaluation-performance-updated}
\end{figure*}

\paragraph{Simplification}

\autoref{fig:evaluation-simplification} shows performance and results for the simplification algorithm described in \autoref{sec:simplification}.
We measured the running time of the algorithm on our benchmarks with simplification enabled (technically, \Algos{L1} with the improvements described above followed by simplification, which we denote as \Algoss{L1}), as well as the total number of corners in the outlines of the unsimplified and simplified layouts.
The final two columns show the total number of corners divided by the total number of rocks.

The results in \autoref{fig:evaluation-simplification} reveal that simplification succeeds in significantly reducing the number of corners in a rock, but is also impractical for large files because it is too slow.
Our largest benchmark, core.ts, which has about three times as many rocks as the second-largest benchmark, did not terminate in a reasonable time, so its corners were not counted.
For simplification to be practical, it needs to be significantly optimized, or used in situations where only a small portion of the document is shown at once.
One property of simplification is useful for the latter case;
because it does not change the position of fragments in the layout, the simplified outlines could be overlaid atop the unsimplified result once they were ready, without causing the document to be re-laid out from scratch.

\begin{figure*}[h]
  \begin{tabular}{ l || r || c || r r || c c }
    &
    &
    \columnAsCaption{Time (s)} 
    &
    \multicolumn{2}{c||}{\columnAsCaption{Total Corners}}
    &
    \multicolumn{2}{c}{\columnAsCaption{Mean Corners}}
    \\
    &
    Rocks
    &
    \rotheader{p{0.0in}}{\Algoss{L1}}
    &
    \multicolumn{1}{c}{\Algos{L1}}
    &
    \multicolumn{1}{c||}{\Algoss{L1}}
    &
    \multicolumn{1}{c}{\Algos{L1}}
    &
    \multicolumn{1}{c}{\Algoss{L1}}
    \\
    \hline
    core.ts   & 8,803 & --- & --- & --- & --- & --- \\
 diff-objs.ts &   106 &               \pz{}\pz{}0.05 & \pz{}3,802 & \pz{}\pz{}\pz{}694 & 36 & 6.5 \\
functional.py & 2,300 &     150\phantom{.}\pz{}\pz{} &     57,074 &             15,574 & 25 & 6.8 \\
   simplex.py &   469 &               \pz{}\pz{}0.97 & \pz{}9,292 &         \pz{}2,358 & 20 & 5.0 \\
     solve.hs & 2,800 & \pz{}93\phantom{.}\pz{}\pz{} &     62,326 &             24,102 & 22 & 8.6 \\
    layout.hs &   865 &           \pz{}\pz{}4.0\pz{} &     16,354 &         \pz{}5,120 & 19 & 5.9 \\
  \end{tabular}
  \caption{Simplification benchmarks.}
  \label{fig:evaluation-simplification}
\end{figure*}

\clearpage
\section{Example Layouts}

\begin{figure}[h]
\includegraphics[width=6.5in]{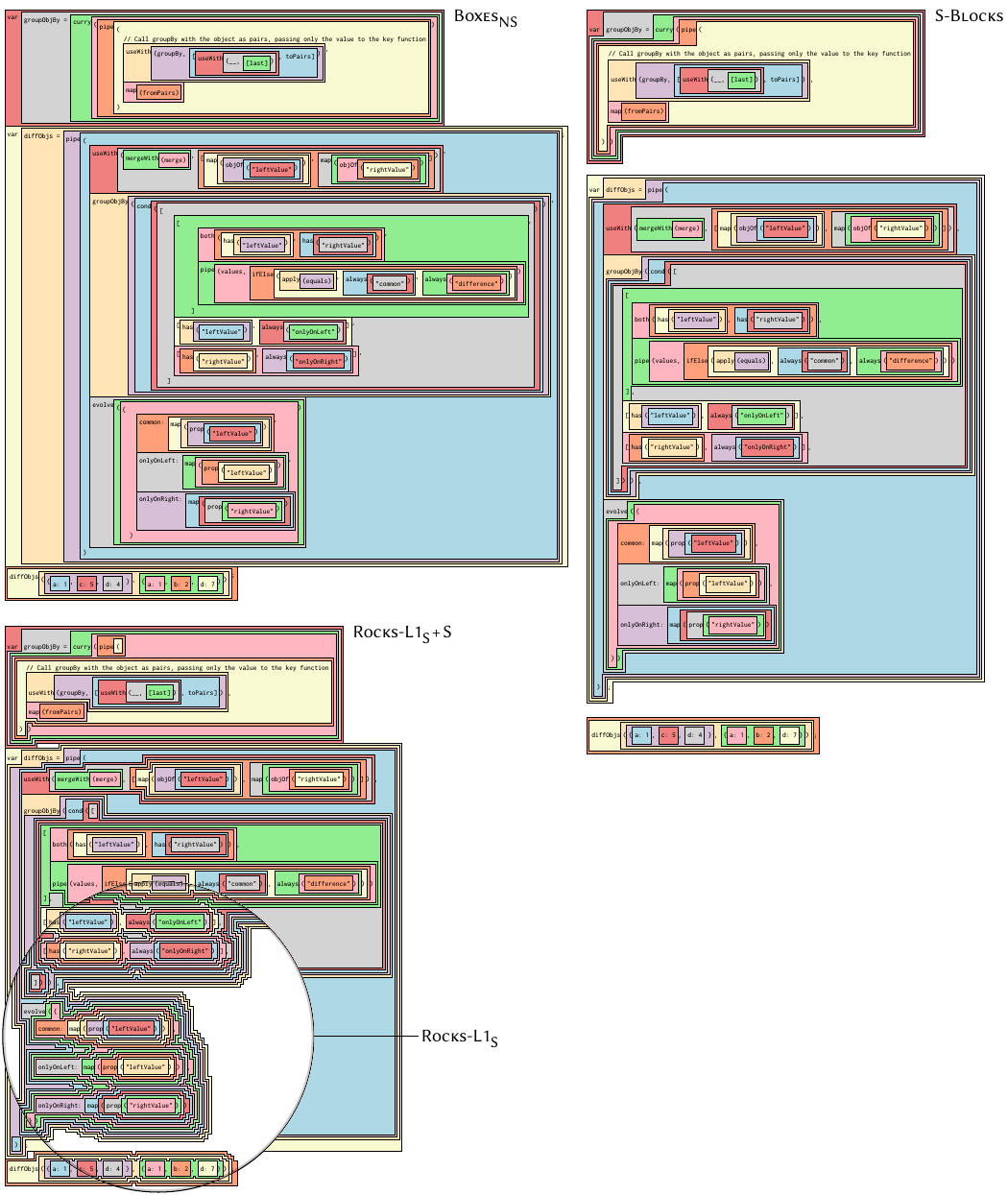} 
\caption{Four layouts for diff-objs with (arbitrarily-colored) blocks everywhere. Notice that (i) the \SBlock{} layout is narrower than \Boxesns{}, (ii) \Algoss{L1} is shorter than \SBlock{}, and (iii) text fragments are identically positioned in \Algoss{L1} and \Algos{L1}.}
\end{figure}

\clearpage
\section{Layout with In-Place Wraps}
\label{sec:version-1i}

\begin{figure}[b]
  \includegraphics[width=3.3in]{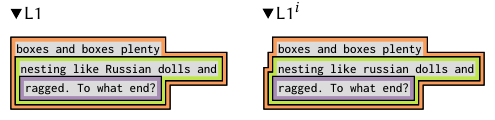} 
  \caption{
A text layout rendered with \Algop{L1} and \Algoi{}.
\Algop{L1} left-aligns the \emph{outermost outline} on each line, while \Algoi{} left-aligns the \emph{first fragment} on each line.
}
  \label{fig:in-place-wraps}
\end{figure}

\newcommand{\termPoint}{$\mathbf{o}$}

\begin{figure}[b]
  \vspace{0.25in} 

  \begin{minipage}{\textwidth}

  \begin{center}
    \begin{tabular}{l c l}
      $\term{layout}$ & \pdn{} & $\List{\termRegionVector{}\hspace{0.01in}}$
      \hspace{0.0in} \text{where} \termRegionVector~::= (\termRegion, \new{\termPoint,} \termVector) \text{and} \new{\termPoint~::= $(\termNum, \termNum)$,} \termVector~::= $\langle\termNum, \termNum\rangle$
    \end{tabular}
  \end{center}

  \beforeNextSectionLine
  \sectionLabel{Union \& Wrap}

  \begin{flalign*}
    \unionl{}~(a, \new{$\mathbf{o}_a,$} \mathbf{v}_a)~(b, \new{$\mathbf{o}_b,$} \mathbf{v}_b) &= (\unionr{}~a~b, \new{$\mathbf{o}_a, (\mathbf{o}_b + \mathbf{v}_b) - \mathbf{o}_a$}) \\[8pt]
    \wrapl{}~\term{id}~\term{padding}~(region, \mathbf{o}, \mathbf{v})
    &=
    (\term{region}',\new{$\mathbf{o} + \langle-\term{padding}, 0\rangle,$} \mathbf{v} + \langle2 \times \term{padding}, 0\rangle) \\
    &\qquad\quad\text{where $\term{region}'=\wrapr{}~\term{id}~\term{padding}~\term{region}$ \removed{translated right by \term{padding}}}
  \end{flalign*}

  \beforeNextSectionLine
  \sectionLabel{Layout}

  \begin{align*}
    \joinh~a~b &=
    \makebox[1.12in][l]{$a~\doubleplus{}~b$}
    \text{when } a = [] \text{ or } b = [] \\
    \joinh~a~b &=
    \makebox[1.12in][l]{$a'~\doubleplus{}~[\termRegionVector\hspace{0.01in}]~\doubleplus{}~b'$}
    \text{otherwise } \\ 
    &\qquad\text{where} \\
    &\qquad\qquad\text{$a'$ (resp. $b'$) are all but the last (resp. first) line of $a$ (resp. $b$),} \\
    &\qquad\qquad\text{\new{$l$ (resp. $r$) is the last (resp. first) line of $a$ (resp. $b$),}} \\
    &\qquad\qquad\text{\new{$l = (\_, \mathbf{o}_l, \mathbf{v}_l)$ and $r = (\_, \mathbf{o}_r, \_)$,}} \\
    &\qquad\qquad\text{$\termRegionVector = \unionl{}~l~r_\text{T}$} \\
    &\qquad\qquad\text{\new{$r_\text{T}$ is $r$ translated by $(\mathbf{o}_l + \mathbf{v}_l) - \mathbf{o}_r$}}
  \end{align*}

  \caption{Algorithm \Algoi{} (extended from \autoref{fig:L1p}).}
  \label{fig:version-1i}
  \end{minipage}
\end{figure}

Our presentation of \Algop{L1} in \autoref{sec:version-1} defines \wrapl{}, the operation which wraps a \emph{layout}, and \emph{translates} the contents of the layout prior to wrapping.
This makes the leftmost edge of the layout stay put under applications of \wrapl{} (the whole layout gets ``bigger'' by \emph{padding}, and is also translated right by \emph{padding}, leaving the position of the left edge unchanged).

But, we could also define \wrapl{} to keep its argument layout ``in place.''
This change complicates some other parts of the layout algorithm (which we'll now refer to as \Algoi{} to disambiguate it from the version of \Algop{L1} which \emph{does} translate in \wrapl{}), but has the potentially desirable effect of making the resulting layouts ever slightly more compact, and maintaining alignment of the first fragment on a line (see \autoref{fig:in-place-wraps}).
The additional complexity arises because the \emph{advance vector} no longer captures enough information about how to horizontally join two \termRegionVector{}s.

To see why, recall that in algorithm \Algop{L1}, when we put one layout (call it $a$) \emph{after} another (call it $b$), we need only translate $b$ by $a$'s \emph{advance}.
(The \emph{advance} of a region is a vector which points from its origin---always $\mathbf{0}$ in \Algop{L1}---to the place where a neighboring region should be attached.)
If \wrapl{} doesn't translate its argument, however, its argument's origin may deviate from $\mathbf{0}$, and so \emph{advance}, while still representing the vector pointing from a region's origin to its attachment point, is not sufficient to reify the position of a horizontally-joined neighbor.
Predictably, we can fix this by associating with each \termRegion{} not only its advance, but also an explicit representation of the region's origin. (See the change to the definition of \emph{layout} in \autoref{fig:version-1i}.)

The definition of \wrapl{} under this new scheme doesn't translate its argument's region, but instead translates its argument's origin.
\unionl{} and \joinh{} are also modified to place their right-hand side at the left-hand side's attachment point, accounting for the possibility that either operand may have a non-zero origin.


\begin{thebibliography}{31}


\ifx \showCODEN    \undefined \def \showCODEN     #1{\unskip}     \fi
\ifx \showISBNx    \undefined \def \showISBNx     #1{\unskip}     \fi
\ifx \showISBNxiii \undefined \def \showISBNxiii  #1{\unskip}     \fi
\ifx \showISSN     \undefined \def \showISSN      #1{\unskip}     \fi
\ifx \showLCCN     \undefined \def \showLCCN      #1{\unskip}     \fi
\ifx \shownote     \undefined \def \shownote      #1{#1}          \fi
\ifx \showarticletitle \undefined \def \showarticletitle #1{#1}   \fi
\ifx \showURL      \undefined \def \showURL       {\relax}        \fi
\providecommand\bibfield[2]{#2}
\providecommand\bibinfo[2]{#2}
\providecommand\natexlab[1]{#1}
\providecommand\showeprint[2][]{arXiv:#2}

\bibitem[Andersen et~al\mbox{.}(2020)]%
        {Andersen2020}
\bibfield{author}{\bibinfo{person}{Leif Andersen}, \bibinfo{person}{Michael
  Ballantyne}, {and} \bibinfo{person}{Matthias Felleisen}.}
  \bibinfo{year}{2020}\natexlab{}.
\newblock \showarticletitle{{Adding Interactive Visual Syntax to Textual
  Code}}.
\newblock \bibinfo{journal}{\emph{Object Oriented Programming Systems Languages
  {\&} Applications {OOPSLA}}} (\bibinfo{year}{2020}).
\newblock
\href{https://doi.org/10.1145/3428290}{doi:\nolinkurl{10.1145/3428290}}


\bibitem[Bar-Yehuda and Ben-Hanoch(1996)]%
        {BarYehuda}
\bibfield{author}{\bibinfo{person}{Reuven Bar-Yehuda} {and}
  \bibinfo{person}{Eyal Ben-Hanoch}.} \bibinfo{year}{1996}\natexlab{}.
\newblock \showarticletitle{{A Linear-Time Algorithm for Covering Simple
  Polygons with Similar Rectangles}}.
\newblock \bibinfo{journal}{\emph{International Journal of Computational
  Geometry \& Applications}} (\bibinfo{year}{1996}).
\newblock
\href{https://doi.org/10.1142/S021819599600006X}{doi:\nolinkurl{10.1142/S021819599600006X}}


\bibitem[Beckmann et~al\mbox{.}(2020)]%
        {Sandblocks2020}
\bibfield{author}{\bibinfo{person}{Tom Beckmann}, \bibinfo{person}{Stefan
  Ramson}, \bibinfo{person}{Patrick Rein}, {and} \bibinfo{person}{Robert
  Hirschfeld}.} \bibinfo{year}{2020}\natexlab{}.
\newblock \showarticletitle{{Visual Design for a Tree-Oriented Projectional
  Editor}}. In \bibinfo{booktitle}{\emph{International Conference on the Art,
  Science, and Engineering of Programming}}.
\newblock
\href{https://doi.org/10.1145/3397537.3397560}{doi:\nolinkurl{10.1145/3397537.3397560}}


\bibitem[Beckmann et~al\mbox{.}(2023)]%
        {Sandblocks2023}
\bibfield{author}{\bibinfo{person}{Tom Beckmann}, \bibinfo{person}{Patrick
  Rein}, \bibinfo{person}{Stefan Ramson}, \bibinfo{person}{Joana Bergsiek},
  {and} \bibinfo{person}{Robert Hirschfeld}.} \bibinfo{year}{2023}\natexlab{}.
\newblock \showarticletitle{{Structured Editing for All: Deriving Usable
  Structured Editors from Grammars}}. In \bibinfo{booktitle}{\emph{{Conference
  on Human Factors in Computing Systems (CHI)}}}.
\newblock
\href{https://doi.org/10.1145/3544548.3580785}{doi:\nolinkurl{10.1145/3544548.3580785}}


\bibitem[Blinn(2019)]%
        {Fructure}
\bibfield{author}{\bibinfo{person}{Andrew Blinn}.}
  \bibinfo{year}{2019}\natexlab{}.
\newblock \bibinfo{title}{{Fructure: A Structured Editing Engine in Racket}}.
\newblock \bibinfo{howpublished}{RacketCon}.
\newblock
\urldef\tempurl%
\url{https://www.youtube.com/watch?v=CnbVCNIh1NA}
\showURL{%
\tempurl}


\bibitem[Burnett(1999)]%
        {BurnettVP}
\bibfield{author}{\bibinfo{person}{Margaret~M. Burnett}.}
  \bibinfo{year}{1999}\natexlab{}.
\newblock \bibinfo{booktitle}{\emph{{Visual Programming}}}.
\newblock \bibinfo{publisher}{John Wiley \& Sons, Ltd}.
\newblock
\href{https://doi.org/10.1002/047134608X.W1707}{doi:\nolinkurl{10.1002/047134608X.W1707}}


\bibitem[Cohen and Chugh(2025a)]%
        {Hass}
\bibfield{author}{\bibinfo{person}{Sam Cohen} {and} \bibinfo{person}{Ravi
  Chugh}.} \bibinfo{year}{2025}\natexlab{a}.
\newblock \showarticletitle{{Code Style Sheets: CSS for Code}}.
\newblock \bibinfo{journal}{\emph{Proceedings of the ACM on Programming
  Languages (PACMPL), Issue OOPSLA1}} (\bibinfo{year}{2025}).
\newblock
\href{https://doi.org/10.1145/3720421}{doi:\nolinkurl{10.1145/3720421}}


\bibitem[Cohen and Chugh(2025b)]%
        {Rocks}
\bibfield{author}{\bibinfo{person}{Sam Cohen} {and} \bibinfo{person}{Ravi
  Chugh}.} \bibinfo{year}{2025}\natexlab{b}.
\newblock \bibinfo{title}{{Ragged Blocks: Rendering Structured Text with Style
  (Extended Version of UIST 2025 Paper)}}.
\newblock
\href{https://doi.org/10.48550/arXiv.2507.06460}{doi:\nolinkurl{10.48550/arXiv.2507.06460}}


\bibitem[Erwig and Meyer(1995)]%
        {Erwig1995}
\bibfield{author}{\bibinfo{person}{Martin Erwig} {and} \bibinfo{person}{Bernd
  Meyer}.} \bibinfo{year}{1995}\natexlab{}.
\newblock \showarticletitle{{Heterogeneous Visual Languages: Integrating Visual
  and Textual Programming}}.
\newblock \bibinfo{journal}{\emph{Symposium on Visual Languages (VL)}}
  (\bibinfo{year}{1995}).
\newblock
\urldef\tempurl%
\url{https://api.semanticscholar.org/CorpusID:17004705}
\showURL{%
\tempurl}


\bibitem[Han et~al\mbox{.}(2020)]%
        {Textlets}
\bibfield{author}{\bibinfo{person}{Han~L. Han}, \bibinfo{person}{Miguel~A.
  Renom}, \bibinfo{person}{Wendy~E. Mackay}, {and} \bibinfo{person}{Michel
  Beaudouin-Lafon}.} \bibinfo{year}{2020}\natexlab{}.
\newblock \showarticletitle{{Textlets: Supporting Constraints and Consistency
  in Text Documents}}. In \bibinfo{booktitle}{\emph{Conference on Human Factors
  in Computing Systems (CHI)}}.
\newblock
\href{https://doi.org/10.1145/3313831.3376804}{doi:\nolinkurl{10.1145/3313831.3376804}}


\bibitem[Harvey et~al\mbox{.}(2013)]%
        {Snap}
\bibfield{author}{\bibinfo{person}{Brian Harvey}, \bibinfo{person}{Daniel~D.
  Garcia}, \bibinfo{person}{Tiffany Barnes}, \bibinfo{person}{Nathaniel
  Titterton}, \bibinfo{person}{Daniel Armendariz}, \bibinfo{person}{Luke
  Segars}, \bibinfo{person}{Eugene Lemon}, \bibinfo{person}{Sean Morris}, {and}
  \bibinfo{person}{Josh Paley}.} \bibinfo{year}{2013}\natexlab{}.
\newblock \showarticletitle{{SNAP! (Build Your Own Blocks)}}. In
  \bibinfo{booktitle}{\emph{Technical Symposium on Computer Science Education
  (SIGCSE TS)}}.
\newblock
\href{https://doi.org/10.1145/2445196.2445507}{doi:\nolinkurl{10.1145/2445196.2445507}}


\bibitem[Hempel et~al\mbox{.}(2018)]%
        {Hempel2018}
\bibfield{author}{\bibinfo{person}{Brian Hempel}, \bibinfo{person}{Justin
  Lubin}, \bibinfo{person}{Grace Lu}, {and} \bibinfo{person}{Ravi Chugh}.}
  \bibinfo{year}{2018}\natexlab{}.
\newblock \showarticletitle{{Deuce: A Lightweight User Interface for Structured
  Editing}}. In \bibinfo{booktitle}{\emph{International Conference on Software
  Engineering (ICSE)}}.
\newblock
\href{https://doi.org/10.1145/3180155.3180165}{doi:\nolinkurl{10.1145/3180155.3180165}}


\bibitem[{JetBrains}(2024)]%
        {MPS}
\bibfield{author}{\bibinfo{person}{{JetBrains}}.}
  \bibinfo{year}{2011--2024}\natexlab{}.
\newblock \bibinfo{title}{{MPS (Meta Programming System)}}.
\newblock
\urldef\tempurl%
\url{https://en.wikipedia.org/wiki/JetBrains_MPS}
\showURL{%
\tempurl}


\bibitem[Knuth and Plass(1981)]%
        {BreakingParagraphs}
\bibfield{author}{\bibinfo{person}{Donald~E. Knuth} {and}
  \bibinfo{person}{Michael~F. Plass}.} \bibinfo{year}{1981}\natexlab{}.
\newblock \showarticletitle{{Breaking Paragraphs into Lines}}.
\newblock \bibinfo{journal}{\emph{Software: Practice and Experience}}
  (\bibinfo{year}{1981}).
\newblock
\href{https://doi.org/10.1002/spe.4380111102}{doi:\nolinkurl{10.1002/spe.4380111102}}


\bibitem[Ko and Myers(2006)]%
        {Ko2006}
\bibfield{author}{\bibinfo{person}{Amy~J. Ko} {and} \bibinfo{person}{Brad~A.
  Myers}.} \bibinfo{year}{2006}\natexlab{}.
\newblock \showarticletitle{{Barista: An Implementation Framework for Enabling
  New Tools, Interaction Techniques and Views in Code Editors}}. In
  \bibinfo{booktitle}{\emph{{Conference on Human Factors in Computing Systems
  (CHI)}}}.
\newblock
\href{https://doi.org/10.1145/1124772.1124831}{doi:\nolinkurl{10.1145/1124772.1124831}}


\bibitem[Mayer et~al\mbox{.}(2015)]%
        {Mayer2015}
\bibfield{author}{\bibinfo{person}{Mika\"{e}l Mayer}, \bibinfo{person}{Gustavo
  Soares}, \bibinfo{person}{Maxim Grechkin}, \bibinfo{person}{Vu Le},
  \bibinfo{person}{Mark Marron}, \bibinfo{person}{Oleksandr Polozov},
  \bibinfo{person}{Rishabh Singh}, \bibinfo{person}{Benjamin Zorn}, {and}
  \bibinfo{person}{Sumit Gulwani}.} \bibinfo{year}{2015}\natexlab{}.
\newblock \showarticletitle{{User Interaction Models for Disambiguation in
  Programming by Example}}. In \bibinfo{booktitle}{\emph{Symposium on User
  Interface Software \& Technology (UIST)}}.
\newblock
\href{https://doi.org/10.1145/2807442.2807459}{doi:\nolinkurl{10.1145/2807442.2807459}}


\bibitem[Miller and Myers(2002)]%
        {LAPIS2002}
\bibfield{author}{\bibinfo{person}{Robert~C. Miller} {and}
  \bibinfo{person}{Brad~A. Myers}.} \bibinfo{year}{2002}\natexlab{}.
\newblock \showarticletitle{{Multiple Selections in Smart Text Editing}}. In
  \bibinfo{booktitle}{\emph{International Conference on Intelligent User
  Interfaces (IUI)}}.
\newblock
\href{https://doi.org/10.1145/502716.502734}{doi:\nolinkurl{10.1145/502716.502734}}


\bibitem[Moon et~al\mbox{.}(2023)]%
        {tylr}
\bibfield{author}{\bibinfo{person}{David Moon}, \bibinfo{person}{Andrew Blinn},
  {and} \bibinfo{person}{Cyrus Omar}.} \bibinfo{year}{2023}\natexlab{}.
\newblock \showarticletitle{{Gradual Structure Editing with Obligations}}. In
  \bibinfo{booktitle}{\emph{Symposium on Visual Languages and Human-Centric
  Computing (VL/HCC)}}.
\newblock
\href{https://doi.org/10.1109/VL-HCC57772.2023.00016}{doi:\nolinkurl{10.1109/VL-HCC57772.2023.00016}}


\bibitem[Mozilla(2025)]%
        {mdnBoxModel}
\bibfield{author}{\bibinfo{person}{Mozilla}.} \bibinfo{year}{2025}\natexlab{}.
\newblock \bibinfo{title}{{Learn CSS: The Box Model}}.
\newblock
\urldef\tempurl%
\url{{
  https://developer.mozilla.org/en-US/docs/Learn/CSS/Building_blocks/The_box_model;
  https://developer.mozilla.org/en-US/docs/Web/CSS/CSS_box_model/Introduction_to_the_CSS_box_model
  https://developer.mozilla.org/en-US/docs/Learn_web_development/Core/Styling_basics/Box_model
  }}
\showURL{%
\tempurl}


\bibitem[{National Instruments}(2024)]%
        {LabVIEW}
\bibfield{author}{\bibinfo{person}{{National Instruments}}.}
  \bibinfo{year}{2024}\natexlab{}.
\newblock \bibinfo{title}{{Laboratory Virtual Instrument Engineering Workbench
  (LabVIEW)}}.
\newblock
\urldef\tempurl%
\url{https://www.ni.com/labview/}
\showURL{%
\tempurl}


\bibitem[Ni et~al\mbox{.}(2021)]%
        {reCode}
\bibfield{author}{\bibinfo{person}{Wode Ni}, \bibinfo{person}{Joshua Sunshine},
  \bibinfo{person}{Vu Le}, \bibinfo{person}{Sumit Gulwani}, {and}
  \bibinfo{person}{Titus Barik}.} \bibinfo{year}{2021}\natexlab{}.
\newblock \showarticletitle{{reCode: A Lightweight Find-and-Replace Interaction
  in the IDE for Transforming Code by Example}}. In
  \bibinfo{booktitle}{\emph{Symposium on User Interface Software and Technology
  (UIST)}}.
\newblock
\href{https://doi.org/10.1145/3472749.3474748}{doi:\nolinkurl{10.1145/3472749.3474748}}


\bibitem[Omar et~al\mbox{.}(2021)]%
        {OmarLivelits}
\bibfield{author}{\bibinfo{person}{Cyrus Omar}, \bibinfo{person}{David Moon},
  \bibinfo{person}{Andrew Blinn}, \bibinfo{person}{Ian Voysey},
  \bibinfo{person}{Nick Collins}, {and} \bibinfo{person}{Ravi Chugh}.}
  \bibinfo{year}{2021}\natexlab{}.
\newblock \showarticletitle{{Filling Typed Holes with Live GUIs}}. In
  \bibinfo{booktitle}{\emph{Programming Language Design and Implementation
  (PLDI)}}.
\newblock
\href{https://doi.org/10.1145/3453483.3454059}{doi:\nolinkurl{10.1145/3453483.3454059}}


\bibitem[Omar et~al\mbox{.}(2012)]%
        {OmarGraphite}
\bibfield{author}{\bibinfo{person}{Cyrus Omar}, \bibinfo{person}{YoungSeok
  Yoon}, \bibinfo{person}{Thomas~D. LaToza}, {and} \bibinfo{person}{Brad~A.
  Myers}.} \bibinfo{year}{2012}\natexlab{}.
\newblock \showarticletitle{{Active Code Completion}}. In
  \bibinfo{booktitle}{\emph{International Conference on Software Engineering
  (ICSE)}}.
\newblock
\href{https://doi.org/10.1109/ICSE.2012.6227133}{doi:\nolinkurl{10.1109/ICSE.2012.6227133}}


\bibitem[Ottmann et~al\mbox{.}(1984)]%
        {RectilinearConvexHull}
\bibfield{author}{\bibinfo{person}{Thomas Ottmann}, \bibinfo{person}{Eljas
  Soisalon-Soininen}, {and} \bibinfo{person}{Derick Wood}.}
  \bibinfo{year}{1984}\natexlab{}.
\newblock \showarticletitle{{On the Definition and Computation of Rectilinear
  Convex Hulls}}.
\newblock \bibinfo{journal}{\emph{Information Sciences}}
  (\bibinfo{year}{1984}).
\newblock
\href{https://doi.org/10.1016/0020-0255(84)90025-2}{doi:\nolinkurl{10.1016/0020-0255(84)90025-2}}


\bibitem[Pollock et~al\mbox{.}(2024)]%
        {Bluefish}
\bibfield{author}{\bibinfo{person}{Josh Pollock}, \bibinfo{person}{Catherine
  Mei}, \bibinfo{person}{Grace Huang}, \bibinfo{person}{Elliot Evans},
  \bibinfo{person}{Daniel Jackson}, {and} \bibinfo{person}{Arvind
  Satyanarayan}.} \bibinfo{year}{2024}\natexlab{}.
\newblock \showarticletitle{{Bluefish: Composing Diagrams with Declarative
  Relations}}. In \bibinfo{booktitle}{\emph{Symposium on User Interface
  Software and Technology (UIST)}}.
\newblock
\href{https://doi.org/10.1145/3654777.3676465}{doi:\nolinkurl{10.1145/3654777.3676465}}

\newpage

\bibitem[Resnick et~al\mbox{.}(2009)]%
        {Resnick2009}
\bibfield{author}{\bibinfo{person}{Mitchel Resnick}, \bibinfo{person}{John
  Maloney}, \bibinfo{person}{Andr{\'e}s Monroy-Hern{\'a}ndez},
  \bibinfo{person}{Natalie Rusk}, \bibinfo{person}{Evelyn Eastmond},
  \bibinfo{person}{Karen Brennan}, \bibinfo{person}{Amon Millner},
  \bibinfo{person}{Eric Rosenbaum}, \bibinfo{person}{Jay Silver},
  \bibinfo{person}{Brian Silverman}, {et~al\mbox{.}}}
  \bibinfo{year}{2009}\natexlab{}.
\newblock \showarticletitle{{Scratch: Programming for All}}.
\newblock \bibinfo{journal}{\emph{Communications of the ACM (CACM)}}
  (\bibinfo{year}{2009}).
\newblock
\href{https://doi.org/10.1145/1592761.1592779}{doi:\nolinkurl{10.1145/1592761.1592779}}


\bibitem[Wu et~al\mbox{.}(2023)]%
        {FFL}
\bibfield{author}{\bibinfo{person}{Zhiyuan Wu}, \bibinfo{person}{Jiening Li},
  \bibinfo{person}{Kevin Ma}, \bibinfo{person}{Hita Kambhamettu}, {and}
  \bibinfo{person}{Andrew Head}.} \bibinfo{year}{2023}\natexlab{}.
\newblock \showarticletitle{{FFL: A Language and Live Runtime for Styling and
  Labeling Typeset Math Formulas}}. In \bibinfo{booktitle}{\emph{Symposium on
  User Interface Software and Technology (UIST)}}.
\newblock
\href{https://doi.org/10.1145/3586183.3606731}{doi:\nolinkurl{10.1145/3586183.3606731}}


\bibitem[Yan et~al\mbox{.}(2024)]%
        {Ivie}
\bibfield{author}{\bibinfo{person}{Litao Yan}, \bibinfo{person}{Alyssa Hwang},
  \bibinfo{person}{Zhiyuan Wu}, {and} \bibinfo{person}{Andrew Head}.}
  \bibinfo{year}{2024}\natexlab{}.
\newblock \showarticletitle{{Ivie: Lightweight Anchored Explanations of
  Just-Generated Code}}. In \bibinfo{booktitle}{\emph{Conference on Human
  Factors in Computing Systems (CHI)}}.
\newblock
\href{https://doi.org/10.1145/3613904.3642239}{doi:\nolinkurl{10.1145/3613904.3642239}}


\bibitem[Ye et~al\mbox{.}(2020)]%
        {Penrose}
\bibfield{author}{\bibinfo{person}{Katherine Ye}, \bibinfo{person}{Wode Ni},
  \bibinfo{person}{Max Krieger}, \bibinfo{person}{Dor Ma’ayan},
  \bibinfo{person}{Jenna Wise}, \bibinfo{person}{Jonathan Aldrich},
  \bibinfo{person}{Joshua Sunshine}, {and} \bibinfo{person}{Keenan Crane}.}
  \bibinfo{year}{2020}\natexlab{}.
\newblock \showarticletitle{{Penrose: From Mathematical Notation to Beautiful
  Diagrams}}.
\newblock \bibinfo{journal}{\emph{ACM Transactions on Graphics}}
  (\bibinfo{year}{2020}).
\newblock
\href{https://doi.org/10.1145/3386569.3392375}{doi:\nolinkurl{10.1145/3386569.3392375}}


\bibitem[Yen et~al\mbox{.}(2024)]%
        {CoLadder}
\bibfield{author}{\bibinfo{person}{Ryan Yen}, \bibinfo{person}{Jiawen Zhu},
  \bibinfo{person}{Sangho Suh}, \bibinfo{person}{Haijun Xia}, {and}
  \bibinfo{person}{Jian Zhao}.} \bibinfo{year}{2024}\natexlab{}.
\newblock \showarticletitle{{CoLadder: Manipulating Code Generation via
  Multi-Level Blocks}}. In \bibinfo{booktitle}{\emph{Symposium on User
  Interface Software and Technology (UIST)}}.
\newblock
\href{https://doi.org/10.1145/3654777.3676357}{doi:\nolinkurl{10.1145/3654777.3676357}}


\bibitem[Yessenov et~al\mbox{.}(2013)]%
        {STEPS}
\bibfield{author}{\bibinfo{person}{Kuat Yessenov}, \bibinfo{person}{Shubham
  Tulsiani}, \bibinfo{person}{Aditya Menon}, \bibinfo{person}{Robert~C.
  Miller}, \bibinfo{person}{Sumit Gulwani}, \bibinfo{person}{Butler Lampson},
  {and} \bibinfo{person}{Adam Kalai}.} \bibinfo{year}{2013}\natexlab{}.
\newblock \showarticletitle{{A Colorful Approach to Text Processing by
  Example}}. In \bibinfo{booktitle}{\emph{Symposium on User Interface Software
  and Technology (UIST)}}.
\newblock
\href{https://doi.org/10.1145/2501988.2502040}{doi:\nolinkurl{10.1145/2501988.2502040}}


\end{thebibliography}
\end{document}